\documentclass[12pt]{article}
\usepackage{jheppub}
\pdfoutput=1
\usepackage{amssymb,amsmath}
\usepackage{epsfig}
\usepackage{epstopdf}
\usepackage{latexsym}
\usepackage{graphicx}
\usepackage{subfigure}
\usepackage{grffile}
\usepackage[mathscr]{euscript}

\usepackage{color}
\usepackage{ifthen}

\makeatletter
\def\@fpheader{\relax}
\makeatother

\newcommand{\red}[1]{\textcolor{red}{#1}}
\newboolean{showToPublic}
\setboolean{showToPublic}{false}
\newcommand{\todocmd}[2]{{#1}{\red{#2}}}
\newcommand{\todo}[2]{\ifthenelse {\boolean{showToPublic}} {\todocmd{#1}{}}{\todocmd{}{#2}}}

\def\cJ{{\cal J}}

\def\cL{{\cal L}}

\def\cL{{\cal L}}

\definecolor{cardinal}{rgb}{0.6,0,0}
\definecolor{darkgreen}{rgb}{0,0.5,0}
\definecolor{golden}{rgb}{0.92, 0.7, 0}
\definecolor{midnight}{rgb}{0, 0, 0.5}
\definecolor{darkblue}{rgb}{0.2, 0, 0.8}

\newcommand{\be}{\begin{equation}}
\newcommand{\ee}{\end{equation}}
\newcommand{\bea}{\begin{eqnarray}}
\newcommand{\eea}{\end{eqnarray}}

\subheader{\begin{flushright}
UTTG-26-15
\end{flushright}}

\title{Holographic Entanglement Entropy in Time Dependent Gauss-Bonnet Gravity}
\author[a,b]{Elena Caceres,}
\author[c]{Manuel Sanchez}
\author[a, d]{and Julio Virrueta}
\affiliation[a]{Facultad de Ciencias, Universidad de Colima, Bernal Diaz del Castillo 340, Colima, M\'exico}
\affiliation[b]{Theory Group, Department of Physics, University of Texas, Austin, TX 78712, USA}
\affiliation[c]{
Instituto de Ciencias Nucleares,
Universidad Nacional Aut\'onoma de M\'exico,\\
Apartado Postal 70-543, M\'exico D.F. 04510, M\'exico}
\affiliation[d]{Department of Physics and Astronomy, 
Stony Brook University, Stony Brook, NY 11794}

\emailAdd{elenac@zippy.ph.utexas.edu}
\emailAdd{jmsm.manuel@gmail.com}
\emailAdd{juliocesar.morenovirrueta@stonybrook.edu}

\abstract{We investigate entanglement entropy in Gauss-Bonnet gravity 
following a global quench. It is  known that in dynamical  scenarios the entanglement entropy probe penetrates the apparent horizon. The goal of  this work is to study how far behind the horizon can the entanglement probe reach in a Gauss-Bonnet theory. We find that the behavior is quite different depending on the sign of the Gauss-Bonnet coupling $\lambda_{GB}$. For $\lambda_{GB}>0$ the behavior of the probes is   just as in Einstein gravity; the probes do not reach the singularity but asymptote to a locus behind the apparent horizon. We calculate the minimum radial position $r_{min}$ reached by the probes and show that for  $\lambda_{GB}>0$ they explore less of the spacetime behind the horizon than in Einstein gravity.  On the other hand, for $\lambda_{GB}<0$ the results are strikingly different; for early times  a new family of solutions appears. These new solutions   reach arbitrarily close to the singularity. We calculate the entanglement entropy for the two family of solutions with $\lambda_{GB} <0$  and find that the ones that reach the singularity are the ones of less entanglement entropy. Thus, for $\lambda_{GB} <0$ the holographic entanglement entropy probes further behind the horizon than in Einstein gravity. In fact, for early times it can explore all the way to the singularity.}

\begin{document}

\maketitle
\flushbottom

\section{Introduction}
The gauge gravity correspondence postulates the equivalence of a gravity 
theory and a (non-gravitational) quantum field theory at the boundary.
For many years this duality has been successfully applied to translate intractable problems in strongly
coupled field theories to manageable calculations in the gravity theory.
However, it should be possible to use the correspondence in the reverse direction and learn about quantum gravity. In particular, it should be possible  to reconstruct the bulk geometry from 
CFT boundary data. Much work has been done in this direction in the last few years and many puzzles and surprises have been 
found along the way. The holographic entanglement entropy prescription (HEE)  of Ryu and Takayanagi (RT) \cite{Ryu:2006bv}  has played a crucial role in these discoveries. Unfortunately, our understanding of how local bulk information is encoded in the boundary  theory is not 
yet complete.

Since AdS/CFT geometrizes field theory observables relating them to geometrical constructions one can ask {\it how much of the 
spacetime geometry is accessible to the  field theory observables?}. This line of work was started  in \cite{AbajoArrastia:2010yt},\cite{Hubeny:2013gta} and developed in \cite{Hubeny:2013dea}. In \cite{AbajoArrastia:2010yt} it was shown that spacelike extremal surfaces cannot penetrate the horizon of an asymptotically  AdS static black hole but they 
do  penetrate the horizon of a dynamical black hole. It was also noted \cite{AbajoArrastia:2010yt} \cite{Hubeny:2013gta} that even though the 
probes can explore behind the horizon they do not reach arbitrarily close to the singularity. This behavior was later shown to be related to the linear growth of the entanglement entropy as a function of time \cite{Liu:2013qca}.

A parallel observation in this line of thought comes from the study of HEE in static AdS black holes with compact boundaries. Typically there are two families of extremal surfaces \cite{Hubeny:2013gta} and we are instructed to choose the one of minimal area as the dual of the entanglement entropy. Which family is the one of minimal area changes depending on where we are in parameter space. 
The switchover  defines a region that the entanglement probe cannot  explore and that is dubbed {\it entanglement shadow }\cite{Freivogel:2014lja} . 

Given these limitations of the HEE to access the bulk and thus be used to reconstruct it, it is natural to consider  that the HEE might
not be the right probe. Indeed, other objects (causal holographic information \cite{Hubeny:2012wa}, entwinement \cite{Balasubramanian:2014sra})  have been proposed either as more natural constructions or as being more suited for bulk reconstruction but it is  still not clear what is their field theory dual. 

In this work we take a different approach; instead of changing the probe we investigate a different gravity theory.  The limitations of the HEE to  access the bulk that we mentioned above have been studied in Einstein gravity. Thus, it is natural to ask how are   these results modified in a  higher derivative theory. Such is the aim of the present work.  We study holographic entanglement entropy in a dynamical (Vaydia type)  scenario in Gauss-Bonnet gravity where the collapse of a null shell results in the formation of an asymptotically AdS black hole. 

Previous studies of HEE in this background have been focused on thermalization time of the field theory. Let us note that in the context of AdS/CFT a Gauss-Bonnet background is dual to a theory with central charges $a\ne c$, and the higher derivative terms arise as corrections in the inverse t'Hooft coupling. The values of $\lambda_{GB}$ were known to be constraint to a small window by demanding causality of the field theory \cite{Buchel:2009sk},\cite{Hofman:2008ar}. These bounds were recently strengthen \cite{Camanho:2014apa} to show that a consistent holographic theory of quantum gravity with finite $\lambda_{GB}$ would require an infinite tower of higher spin fields. However, it is still possible to treat the higher derivative terms as part of a perturbative  expansion of a string or quantum gravity theory. Thus, such bottom-up approach is used in the thermalization time \cite{Li:2013cja}  \cite{Zeng:2013mca} \cite{Zeng:2013fsa} and anisotropic plasmas in Gauss-Bonnet \cite{Jahnke:2014vwa}.

Our interest is not to make contact with a field theory but is more geometric, we want to understand if in  higher derivative theories HEE can probe further behind the horizon than in Einstein gravity. In \cite{Hubeny:2007xt} Hubeny, Rangamani and Takayanagi (HRT) presented a
covariant generalization of the holographic entanglement entropy prescription of RT. This proposal enabled the holographic study of time dependent phenomena and out of equilibrium physics. The linear growth of entaglement entropy as a function of time typical of ergodic systems was obtained holographically and the thermalization time explored in scenarios with and without chemical potential\cite{Balasubramanian:2011ur},\cite{Caceres:2012em}, \cite{
Galante:2012pv}, \cite{Caceres:2012px}.
The extension of the RT proposal to Gauss-Bonnet theories was initiated in \cite{Hung:2011xb} and \cite{deBoer:2011wk}. In \cite{Dong:2013qoa},\cite{Camps:2013zua} the authors generalized the proposal  to arbitrary higher derivatives theories and presented a covariant prescription. Recently, in \cite{Headrick:2014cta} it was shown that as long as long as the bulk obeys the null energy condition, the covariant prescription for the entanglement entropy HRT is compatible with causality of the field theory. This point  was further explored in higher derivative theories in \cite{Erdmenger:2014tba}.

In this paper we use these results to study HEE in a Gauss-Bonnet black hole formed by collapsing a null shell. We find that  for $\lambda_{GB}>0$ the HEE surfaces behave just as in Einstein gravity: they penetrate the horizon but stop at a limiting locus  and do not reach the singularity. Furthermore, we calculate the minimum point the geodesics can reach behind the horizon and show that for $\lambda_{GB}>0$ the HEE explores less than HEE in Einstein gravity. For $\lambda_{GB}<0$ the results are strikingly different: at early times the solutions become double valued with one family reaching the singularity and the other not. Given these two
families of solutions the prescription instruct us to choose the one of minimal entropy. We find that it is the family that reaches the singularity the one that has minimal entropy. Thus, our results indicate that for $\lambda_{GB}<0$ the holographic entanglement entropy can explore all the way to the singularity. For later times the two families join and again 
and HEE no longer explores close to the singularity. 

This paper is organized as follows: in Section~\ref{sec:GB} we summarize some known facts of GB gravity and present the background we will use in the rest of the paper. As a warmup we calculate spacelike geodesics in Section~\ref{sec:geo} ans show that already for spacelike geodesics the curves that reach arbitrarily close to the singularity exist when $\lambda_{GB}<0$. In Section~\ref{sec:main} we present the study of  HEE in a Vaidya Gauss-Bonnet background for postive and negative coupling, we explain the numerical procedure and present the results. In the last section (Section~\ref{sec:conclusions}) we present the conclusions and several possible directions that are, in our opinion, interesting to investigate.

\section{Gauss-Bonnet gravity}\label{sec:GB}
In the framework of the AdS/CFT correspondence,  higher derivatives
terms  are expected to arise  as quantum or stringy
corrections to the classical action. Thus, it is compelling to  consider  an effective action where the cosmological constant and Einstein terms are supplemented
by  curvature corrections. In this section we will gather some well known facts about one particular such theory: Gauss-Bonnet gravity.

Let us consider  five dimensional  Gauss-Bonnet gravity. This theory is the simplest of Lovelock theories which are known to yield second order equations of motion in spite containing  higher derivative terms in the action. They are  free of pathologies and are solvable. In fact, many black hole solutions with AdS asymptotics are known \cite{Camanho:2011rj}.

The action is  given by (following the notations in \cite{Myers:2012ed}, see also \cite{deBoer:2011wk})

\begin{align}\label{eq:action}
	&S_{grav}=\frac{1}{16 \pi G_N}\int d^5x \sqrt{-g}\left( R + \frac{12}{L^2}+ \frac{\lambda_{GB} L^2}{2} \mathcal{L}_{(2)}\right),\\
 &\mathcal{L}_{(2)}= R_{\mu \nu\rho\sigma}R^{\mu\nu\rho\sigma} - 4 R_{\mu\nu}R^{\mu\nu} + R^2
\end{align}
Here $G_N$ denotes the five-dimensional Newtons's constant, R denotes the Ricci-scalar; the cosmological constant is given by $\Lambda=-12/L^2$, where $L$ is some length scale. Varying the action in \eqref{eq:action}, we get the following equation of motion in \eqref{eq:action}

\begin{align}
	&R_{\mu\nu}- \frac{1}{2} g_{\mu\nu} \left( R + \frac{12}{L^2} + \frac{\lambda_{GB} L^2}{2}\mathcal{L}_{(2)}\right) + \mathcal{H}_{\mu\nu}^{(2)}=0,\label{eq:Rmunu}\\
&\mathcal{H}_{\mu\nu}^{(2)}= R_{\mu\nu\rho\sigma\lambda_{GB}}R_\nu^{\rho\sigma\lambda_{GB}} - 2 R_{\mu\rho}R_\nu^\rho - 2 R_{\mu\rho\nu\sigma}R^{\rho\sigma} + R R_{\mu\nu}.
\end{align}
In some sense, the tensor $\mathcal{H}_{\mu\nu}^{(2)}$ can be thought of as an external energy-momentum tensor sourced by the higher derivative terms.
 
A solution of the equation of motion \eqref{eq:Rmunu} is\footnote{We do not consider the solution with  $ f(z)=\frac{1}{2\lambda_{GB}}[1+\sqrt{1-4\lambda_{GB}(1-mz^4)}]$ since it contains ghosts and is unstable \cite{Boulware:1985wk} \cite{Myers:2010ru}.}\ \cite{Cai:2001dz},

\begin{align}\label{eq:staticmetric}
	&ds^2= -\frac{L^2}{z^2} \frac{f(z)}{f_0} dv^2 + \frac{L^2}{z^2}\left(-\frac{2}{\sqrt{f_0}} dz dv + d\bar{x}^2\right),\\
	&f(z)=\frac{1}{2\lambda_{GB}}[1-\sqrt{1-4\lambda_{GB}(1-mz^4)}].
\end{align}
The background in \eqref{eq:staticmetric} represents an asymptotically AdS-space black hole solution of Gauss-Bonnet gravity where 
$L$ is related to the radius of curvature and the event-horizon is located at $z_{eh} = m^{-1/4}$. We have expressed the above solution in  Eddington-Finkelstein coordinates, which  are defined
by
\begin{equation}
	dt=dv+ \sqrt{f_0}\frac{dz}{f(z)}
\end{equation}
where $t$ is  the boundary time. Near the boundary, where $z\rightarrow0$, $v\rightarrow t$. The constant $f_0$ has been chosen such that lim$_{z\rightarrow0} f(z)=f_0$, 
\begin{align}
f_0=\frac{1}{2\lambda_{GB}}\left( 1- \sqrt{1- 4\lambda_{GB}}\right).
\end{align}
we chose to normalize the coordinates in \eqref{eq:staticmetric} such that at the boundary $g_{tt}/g_{xx}|_{z\rightarrow 0} = -1$. 
In Poincar\`e patch, the background in \eqref{eq:staticmetric} takes the following form
\begin{align}\label{eq:metricPoincare}
	ds^2= -\frac{L^2}{z^2}\frac{f(z)}{f_0} dt^2 + \frac{L^2}{z^2} d\bar{x}^2 + \frac{L^2}{z^2} \frac{dz^2}{f(z)}.
	\end{align}
Note that the parameter $L$ is related to the $AdS$ curvature scale $\tilde{L}$ as $\tilde{L}^2= L^2/f_0$.
	The temperature of this solution is,
	
	\begin{align}
	&T=\frac{m^{1/4}}{\pi L^2}\frac{1}{\sqrt{f_0}}\nonumber\\
	\end{align}
Note that  the event horizon is always located at 
\begin{equation}\label{eq:z_ev_hor}
z_{eh}= m^{-1/4}
\end{equation}
regardless of the value of $\lambda_{GB}$. The curvature singularity that occurs at $z=\infty$ for $\lambda_{GB}\ge 0$ is shifted to a finite radial position,
\begin{equation}\label{eq:z_sing}
z_s=\frac{1}{\sqrt{2} m^{1/4} \lambda_{GB}^{1/4}}(-1 + 4\lambda_{GB})^{1/4}
\end{equation}
for $\lambda_{GB}<0$. 

Finally, througout this paper we will  work with very small values  of the Gauss-Bonnet coupling, $\lambda_{GB}=\pm0.05$ as representativs of positive and negative Gauss-Bonnet couplings.

\subsection{Time dependent background}

	We will now  us discuss a time-dependent generalization of the background in \eqref{eq:staticmetric}. In order to do so, we need to couple the action $S_{grav}$ in \eqref{eq:action} with an external source term $S_{ext}$ to yield
$$S=S_{grav} + \kappa S_{ext},$$
where  $\kappa$ is some coupling which we do not specify here. A simple solution of the following form can be obtained 
\begin{align}\label{eq:timemetric}
	&ds^2=-\frac{L^2}{z^2} \frac{f(z,v)}{f_0} dv^2 + \frac{L^2}{z^2} \left( -\frac{2}{\sqrt{f_0}} dzdv + d{\bar x}^2 \right), \qquad f_0=\frac{1}{2\lambda_{GB}}( 1-\sqrt{1-4\lambda_{GB}}),\\
	&f(z,v)=\frac{1}{2\lambda_{GB}}\left[ 1 - \sqrt{1-4\lambda_{GB}(1-m(v)z^4)}\right].
\end{align}
Here $m(v)$ is a function that is hitherto undetermined. It is straightforward to check that the external source must yield the following energy-momentum tensor
\begin{equation}\label{eq:tmunu}
	(16 \pi G_N)\kappa T^{ext}_{\mu\nu} = \frac{3}{2} z^3 \frac{dm}{dv} \delta_{\mu v}\delta_{\nu v}.
\end{equation}
Thus a null energy condition on the external energy momentum tensor will give the condition $m'(v)\ge0$. Null energy condition in the bulk is related to strong subadditivity in the boundary\cite{Callan:2012ip},\cite{Caceres:2013dma}. We want to preserve both of them so we will choose a profile that satisfies $m'(v)\ge0$\footnote{This implies that our solutions should produce an $S(\ell)$ that is concave and monotonically increasing \cite{Callan:2012ip} and we will see in Section \ref{subsec:results} that they indeed do.} .
Since this is a time-dependent geometry we need to identify the  apparent horizon.

A trapped surface $T$ is  a co-dimension two spacelike submanifold such that  the expansion of both ``ingoing" and ``outgoing" future directed null geodesics orthogonal to $T$ is everywhere negative. The boundary of the trapped surfaces is the apparent horizon.

\begin{figure}
\centering
\includegraphics[width=3in]{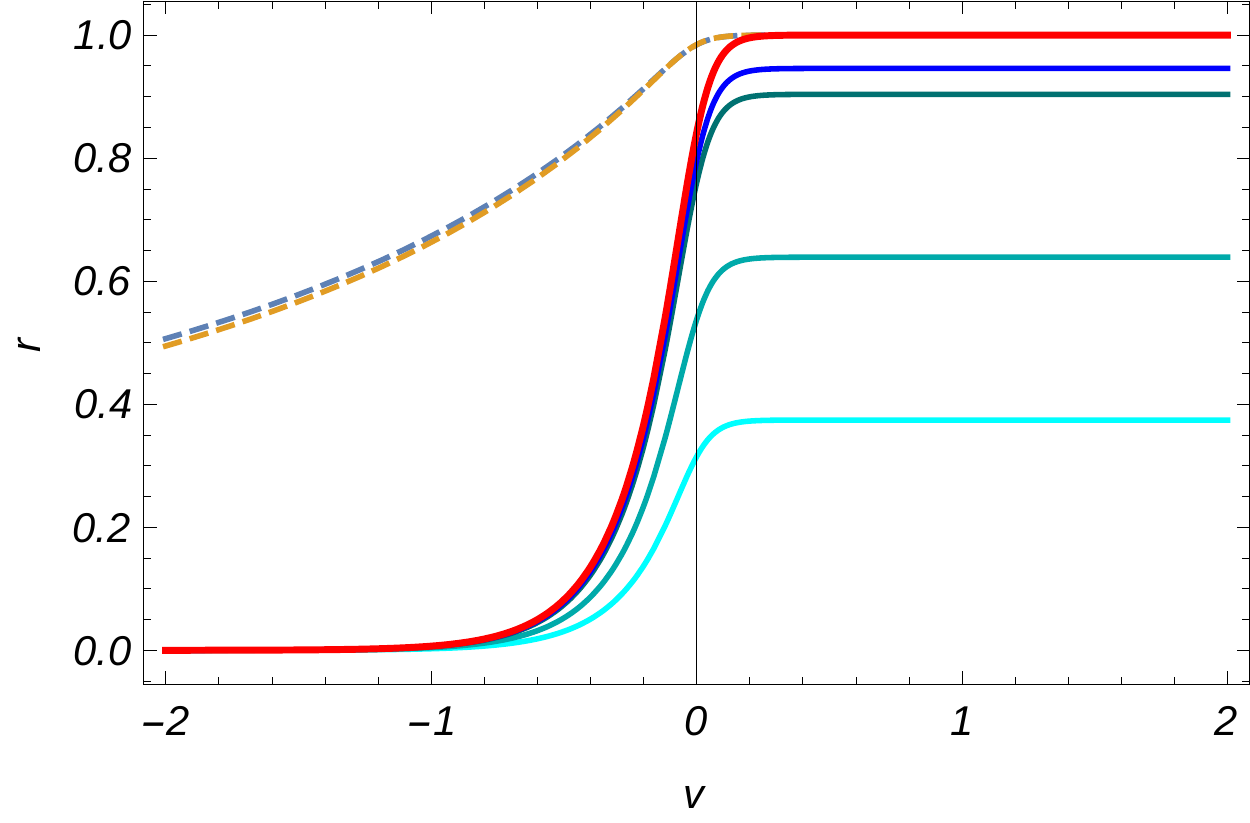}
\caption{The  dashed lines are the event horizons for $\lambda_{GB}= 0.05,\ -0.05$ The  red line  is the apparent horizon which is independent of $\lambda_{GB}$ and the other lines are  the singularity for different values of $\lambda_{GB}=-.0.005,\ -0.05,\ -.5,\ -1.$ from bottom to top respectively. \label{fig:apparent_event_horizons}}
\end{figure}

In what follows, we will closely follow \cite{Figueras:2009iu}. For the background in (\ref{eq:timemetric}) the vectors tangent to the ingoing and outgoing null geodesics are given by
\begin{eqnarray}
l_- = - \partial_z \ , \quad l_+ = - \frac{z^2}{L^2} \partial_v + \frac{z^2}{2 L^2} f \sqrt{f_0} \partial_z \ 
\end{eqnarray}
such that
\begin{eqnarray}
l_- \cdot l_- = 0 \ , \quad l_+ \cdot l_+ = 0 \ , \quad l_- \cdot l_+ = -1 \ .
\end{eqnarray}
Now the volume element of the co-dimension two spacelike surface (orthogonal to the above null geodesics) is given by
\begin{eqnarray}
\Sigma = \left(\frac{L}{z}\right)^3 \ .
\end{eqnarray}
The expansions are defined to be
\begin{eqnarray}
\theta_{\pm} = \cL_{\pm} \log \Sigma = l_{\pm}^\mu \partial_\mu \left( \log\Sigma \right) \ ,
\end{eqnarray}
where $\cL_{\pm}$ denotes the Lie derivatives along the null vectors $l_{\pm}$. The apparent horizon is then obtained by solving the equation $\Theta = 0$, where $\Theta = \theta_+ \theta_-$ is the invariant quantity. In this case we find
\begin{eqnarray} \label{eq:apparent_horizon}
\Theta = \frac{9}{2 L^2} f(z_{ah},v) = 0 
\end{eqnarray}
gives the location of the apparent horizon,
\begin{equation}
z_{ah}=m(v)^{-1/4}
\end{equation}
The event horizon, on the other hand, is a null surface in the geometry \eqref{eq:timemetric},
\begin{equation}
\mathcal{N}=z-z_{eh}(v), \qquad \qquad G^{\mu\nu}\partial_\mu\mathcal{N}\partial_\nu\mathcal{N}=0,
\end{equation}\label{eq:event_horizon}
which gives the evolution of the event horizon,
\begin{equation}
z_{eh}'(v)=-\frac{1}{2 \sqrt{f_0}}f(z_{eh},v).
\end{equation}
It is clear form \eqref{eq:apparent_horizon} that the position of the apparent horizon does not depend on $\lambda_{GB}$. However, the position of the event horizon does albeit mildly (see Figure \ref{fig:apparent_event_horizons})

Our goal in the next sections is to study non-local probes in \eqref{eq:timemetric}  that evolves between a vacuum  AdS geometry with radius and an asymptotically AdS black hole solution of Gauss Bonnet action \eqref{eq:action}. 
We will be particularly interested in comparing how far behind the apparent horizon the different probes can reach. Our main interest is the behavior of the  entanglement entropy, Section \ref{sec:main}. However, we will first analyze 
spacelike geodesics  which  already illustrate some novel features also present in the entanglement entropy. 

\section{Spacelike geodesics}\label{sec:geo}
To find  spacelike geodesics in the background  \eqref{eq:timemetric} we extremize the following Lagrangian

\begin{equation}\label{eq:Lag_geo}
\mathcal{L}=\frac{1}{z(x)} \sqrt{f_0 -f(z,v) v'(x)^2 -2 \sqrt{f_0} v'(x)z'(x)}
\end{equation}

The equations of motion are\footnote{The explicit form of the functionals $\mathcal{Z}[v(x),z(x)]$ and
$\mathcal{V}[v(x),z(x)]$ are given in Appendix \ref{app:eom}}
\begin{align}\label{eq:eom_geo}
&z''(x)= \mathcal{Z}[v(x),z(x)]\nonumber\\
&v''(x)=\mathcal{V}[v(x),z(x)]
\end{align}

The equations of motion will be solved subject to boundary conditions

$$z(l/2)=z(-l/2)= z_0\qquad \qquad \textrm{ and} \qquad\qquad z'(0)=v'(0)=0$$ where  $l/2$ denotes the distance in the boundary and $z_0$ is the IR cutoff. The Lagrangian \eqref{eq:Lag_geo} does not depend explicitly on $x$, the corresponding   conserved quantity is,
\begin{equation}\label{eq:conserved_C_1}
\frac{1}{z(x)^2 \mathcal{L}} = C_1.
\end{equation}

We will be working with a thin shell,  $M(v)=\frac{1}{2}(1+\tanh(v/v_0))$ with $v_0= 0.01$. As the parameter $v_0$ goes to zero, this function approximates a step function. Thus,  in the limit $v_0\rightarrow 0$ the spacetime is given by the gluing of two static geometries.  Let us pause and,  to gain some intuition, briefly study  the behavior of geodesics in  static Gauss Bonnet spaces. 

If $M(v)=m$, $v$ is a  cyclic variable and its associated  momentum  is conserved,
\begin{align}\label{eq:conserved_C_2}
\frac{-f(z)v'(x) - \sqrt{f_0}z'(x)}{z(x)^2 \mathcal{L}} = C_2
\end{align}

Thus, in the static case we have two conserved quantities \eqref{eq:conserved_C_1} and \eqref{eq:conserved_C_2}. It is instructive to find the effective potential in this case since it will illustrate the differences for $\lambda_{GB}>0$ and $\lambda_{GB}<0$ that will persist in the dynamic case. Solving  \eqref{eq:conserved_C_1} for $v'(x)$ and substituting in \eqref{eq:conserved_C_2} we get

\begin{equation}
z'(x)^2 = E^2 -V_{eff}
\end{equation}
where
\begin{align}\label{eq:veff}
V_{eff}&=f(z,\lambda_{GB})( 1-\frac{\cJ^2}{z^2}). \
\end{align}
\begin{figure}
\centering
\mbox{\subfigure{\includegraphics[width=3in]{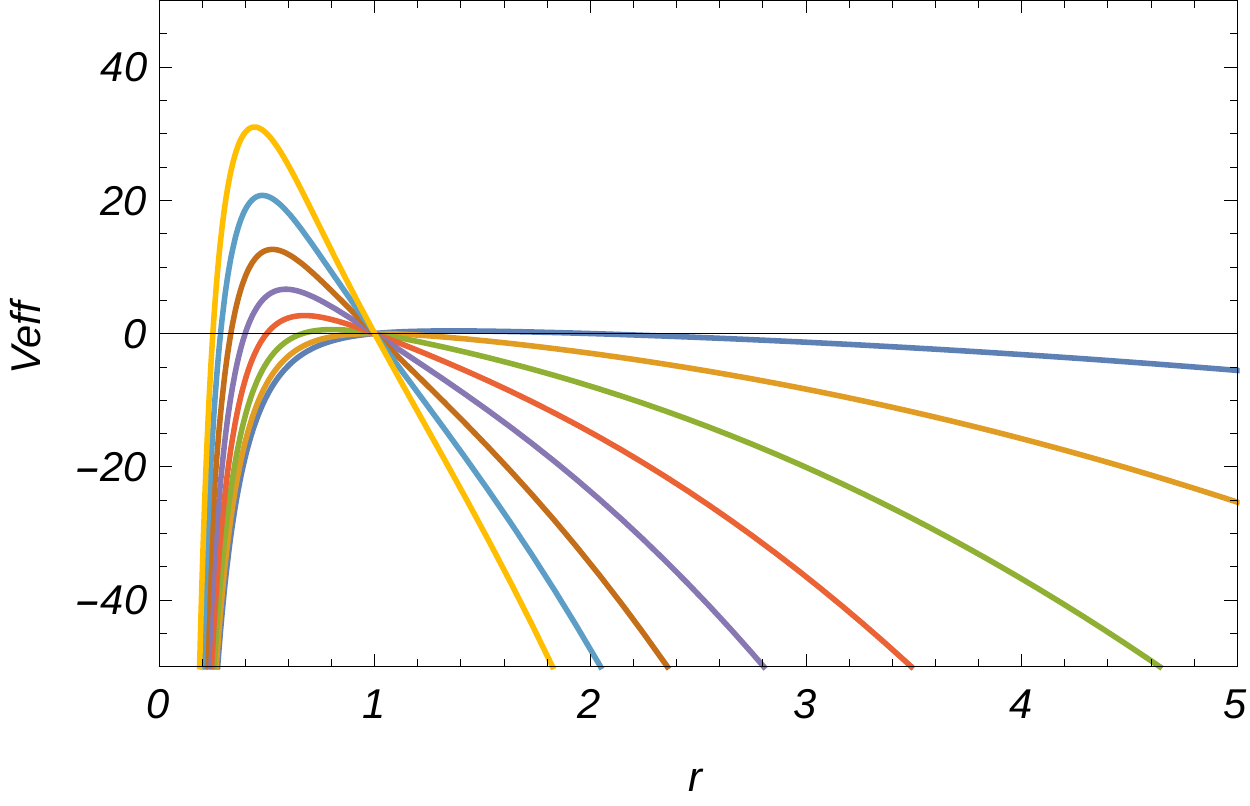}}\quad
\subfigure{\includegraphics[width=3in]{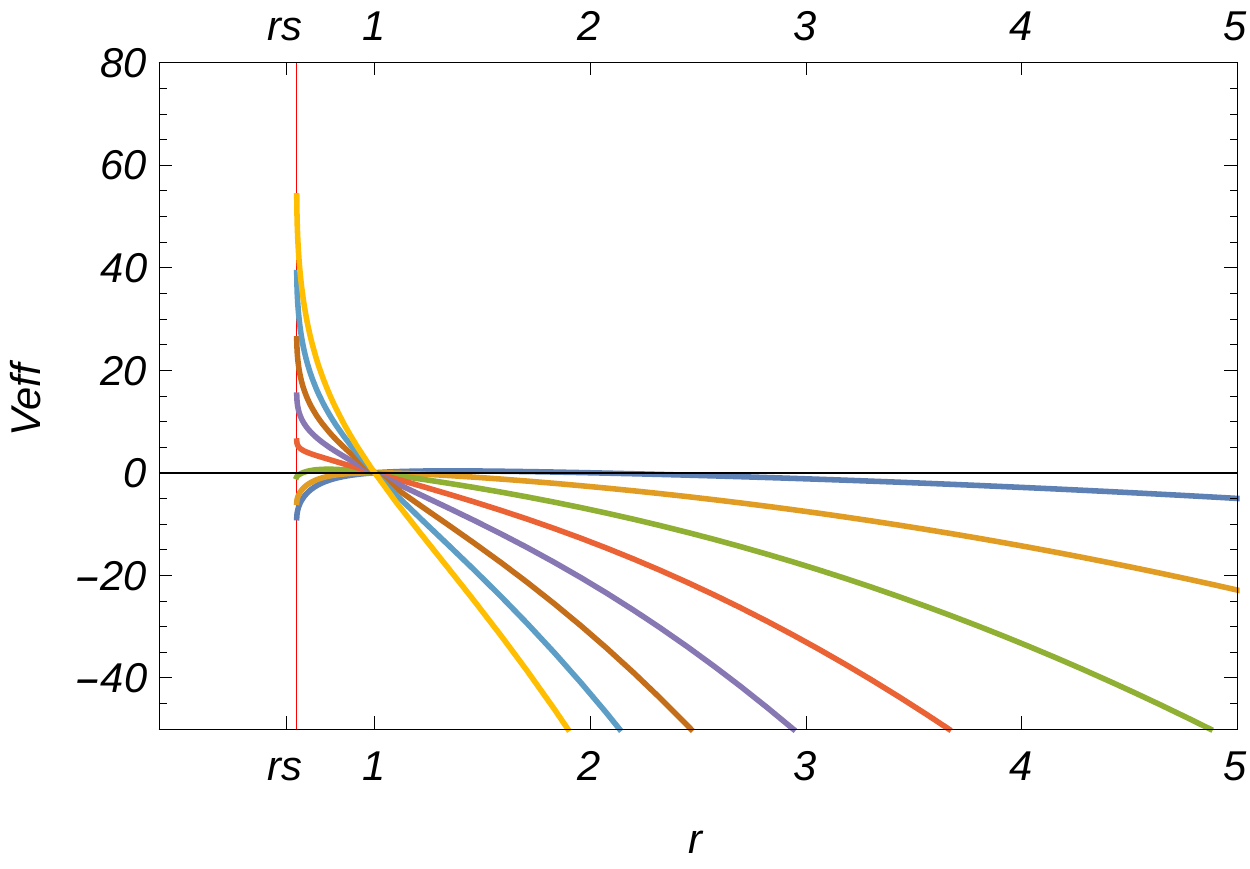} }}
\mbox{\includegraphics[width=3in]{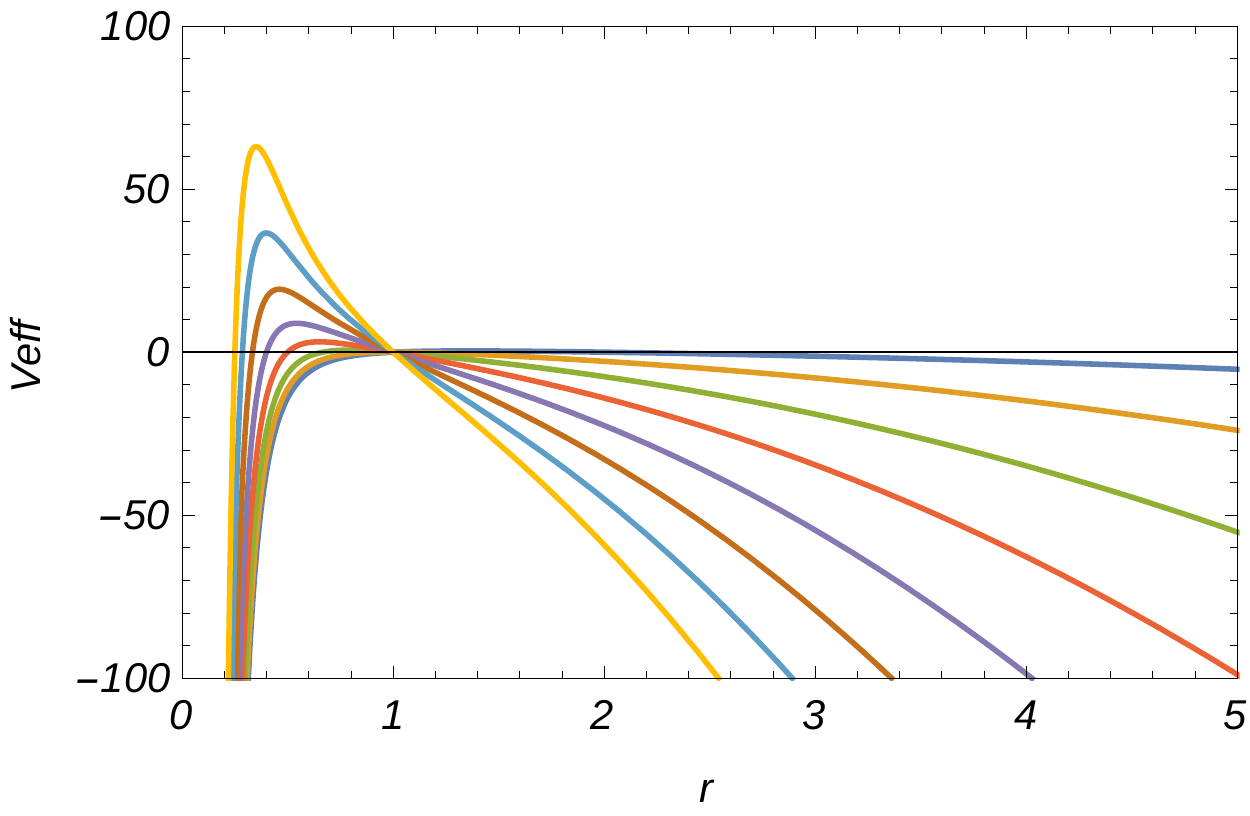}}
\caption{Effective potential $V_{eff}(r)$ for $m=1$ and  different values of $\cJ$. Left panel : $\lambda_{GB}=.05$. \quad Right panel: $\lambda_{GB}=-0.05.$ In the lower panel we have included the $SAdS$ case for comparison.} \label{fig:potential_pos_neg_lambda}
\end{figure}

For convenience we have redefined the constants as $E=\frac{C_2}{C_1}\sqrt{ f_0(\lambda_{GB})}$ and $\cJ=\frac{1}{C_1}$. In Figure (\ref{fig:potential_pos_neg_lambda})  we plot $V_{eff}$ for different values of $C_1$ as a function of $r=1/z$. Note the different behavior of the potential for $\lambda_{GB} >0$ and $\lambda_{GB}<0$. For $\lambda_{GB}>0$  the potential is similar to the Schwarzchild $AdS_5$ ($SAdS_5$); it reaches a maximum for some value of $0<r<rh$ and the value of the maximum grows with large $\cJ$. The growth of this maximum becomes less pronounced as we increase $\lambda_{GB}$. For $\lambda_{GB} <0$ the potential is qualitatively different. For small values of $\cJ$ it is a concave function that reaches its maximum at some $r_s<r< r_h$ similar to the positive $\lambda_{GB}$ case. However, for large $\cJ$ the concavity of the potential changes and reaches its maximum at the singularity $r_s$. We identify the critical value of $\cJ$ at which this change occurs,

\begin{equation}\label{eq:L_crit}
\cJ_{crit}\sim \frac{1}{2\sqrt{-\lambda_{GB}}(1+2\lambda_{GB})}.
\end{equation}
The existence of this different regimes in the case of negative $\lambda_{GB}$ is clearly  asociated with the fact that the singularity has shifted from $r=0$ to $r_s^{-1}=\frac{1}{\sqrt{2} m^{1/4} \lambda_{GB}^{1/4}}(-1 + 4\lambda_{GB})^{1/4}$.
This different behavior of $V_{eff}$ for  $\lambda_{GB}<0$ and $\lambda_{GB}>0$ will be reflected in the time dependent case. 

Now we are ready to proceed with the time dependent case with $M(v)$ in \eqref{eq:timemetric} given by $M(v)=\frac{1}{2}(1+ \tanh{\frac{v}{v_0}})$.
We want to solve the  differential equations \ref{eq:eom_geo} subject to the following boundary conditions
$$z(0) = z_* ,\quad z'(0)=0,\quad
 v(0) = v_* ,\quad v'(0)=0$$
So far $z_*$ and $v_*$ are two free
parameters that generate the numerical solutions for $z(x)$ and $v(x)$. Once a solution is obtained the boundary data can be read off,
$$z(l/2)=z_0\qquad\qquad v(l/2)=t_b$$. 
Note that the numerically nontrivial part is to look for appropriate parameters $(z_*,v_*)$ such that the IR cutoff $z_0$ is a small number \footnote{In our solutions we demand $z_0 \sim 10^{-5}$}. We will expand on the details of the numerical procedure in section (\ref{sec:Numerics}) when we deal with the HEE probes that
 is our main objective.

\begin{figure}[t!]
\centering
\parbox{3in}{\includegraphics[width=3in]{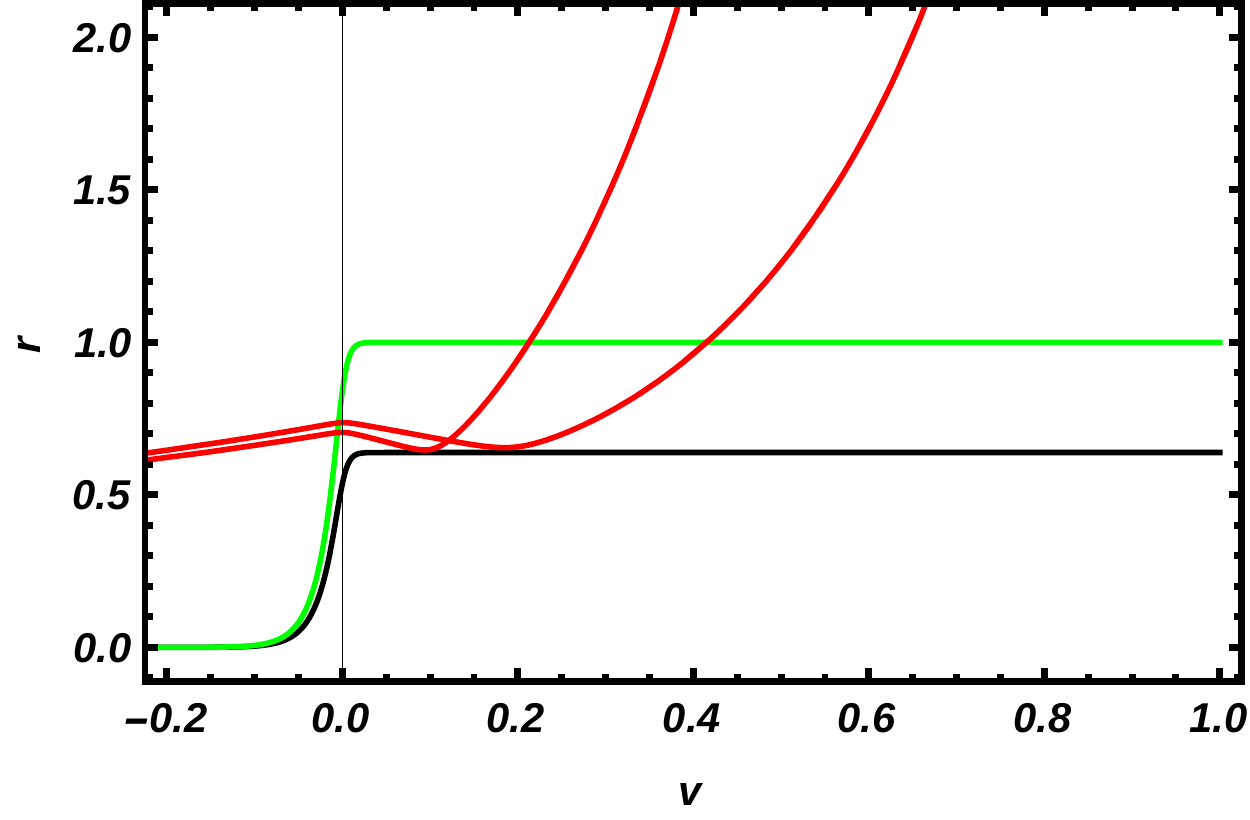}}\quad
\parbox{3in}{\includegraphics[width=3in]{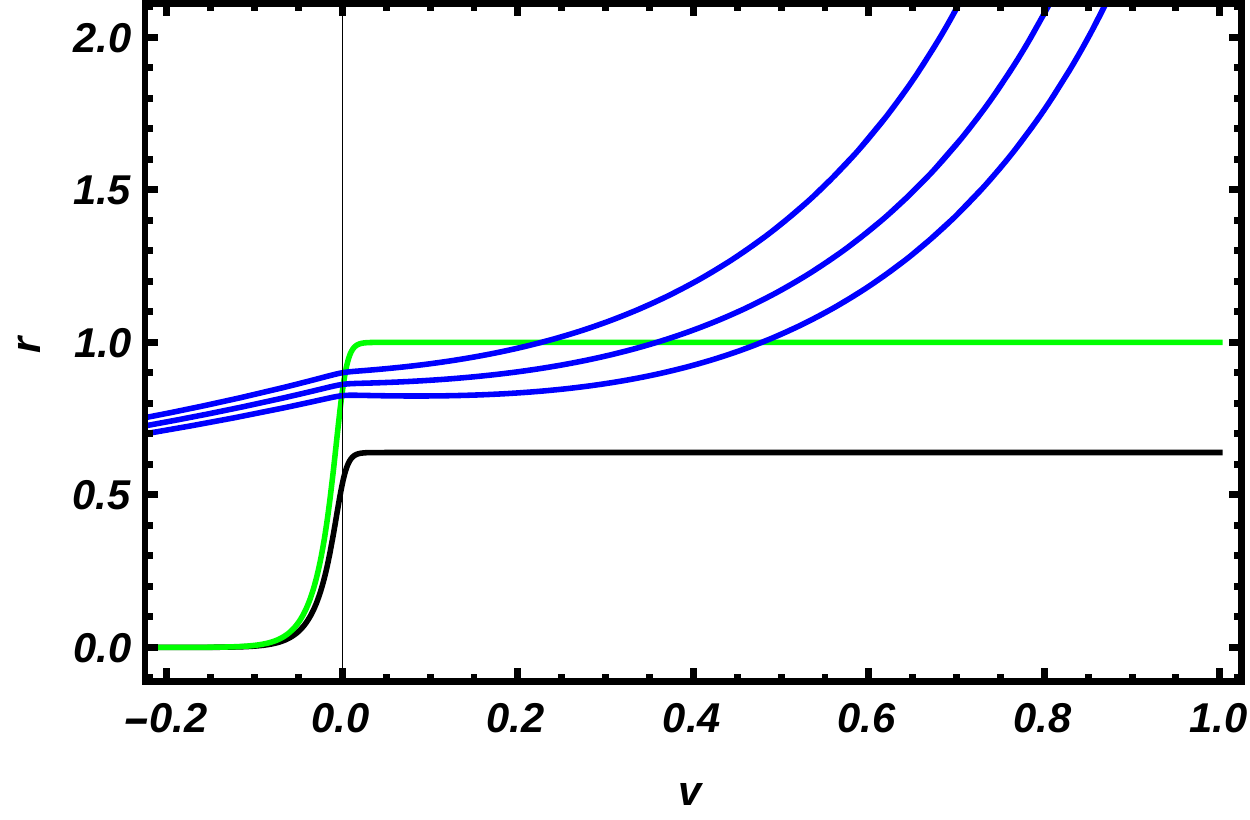} }
\caption{Representative geodesics for negative $\lambda_{GB}$. Left panel: the new family of solutions that reach  the singularity. Right panel: the standard solutions that asymptote to a critical surface and do not reach the singularity.  \label{fig:new_old_geo_negative_lambda}}
\end{figure}

We solve the equations of motion and look for geodesics that cross the horizon and are anchored at the boundary. We find that the case of negative $\lambda_{GB}$ presents some new features. There are geodesics that, as in SAdS-Vaidya, cross the horizon but do not reach the singularity and asymptote to a limiting surface. But there is also -- for some range of parameters-- a new family of boundary anchored geodesics  that do reach arbitrarily close to the singularity. It is remarkable that this strikingly different behavior depends crucially on the sign of $\lambda_{GB}$ and  is present even for such a small value of the coupling as $\lambda_{GB}= -0.05$. 
In Figure (\ref{fig:new_old_geo_negative_lambda}) we plot some representative solutions.

\section{Entanglement Entropy}\label{sec:main}
	The proposal for computing entanglement entropy is modified acording to the formula obtained in \cite{deBoer:2011wk} \cite{Hung:2011xb}

	\begin{align}\label{eq:ee}
		S_{EE}=\frac{1}{4G_N} \int_\Sigma d^3 \xi\sqrt{\gamma}(1+\lambda_{GB} L^2 R_\Sigma) + \frac{1}{2G_N} \int_{\partial \Sigma} d^2 \xi \sqrt{h} K,
	\end{align} 

	\noindent where $\Sigma$ denotes the three dimensional surface anchored at the two dimensional boundary of the region A; $\xi$  denotes the worldvolume coordinates on the surface and $h$ denotes the induced metric on this surface. The quantity $R_\Sigma$  denotes the Ricci scalar constructed from the induced metric on $\Sigma$ and the last term in (\ref{eq:ee}) is the Gibbons-Hawking boundary term that one needs to introduce to have well a defined variational problem. The proposal  \eqref{eq:ee} was initially put forward for time independent situations. In \cite{Dong:2013qoa},\cite{Camps:2013zua} the authors generalize the proposal of \cite{deBoer:2011wk}, \cite{Hung:2011xb} and presented a covariant  prescription valid for more general theories of higher derivatives. 

Throughout this paper we will study the  ``rectangular strip" in the backgrounds \eqref{eq:timemetric}.  Assuming translational invariance in two of the directions we can parametrize the extremal area surface by the third coordinate, $x$, and we have ${z(x), v(x)}$ . We denote the width of the rectangular strip $\ell$, that is $\frac{\ell}{2}< x < \frac{\ell}{2}$. The induced metric on the co-dimension two surface is given by
\begin{equation}\label{eq:codimtwo}
	ds^2=\frac{L^2}{z^2}(dx_2^2 + dx_3^2) + \frac{L^2}{z^2}\left(1- \frac{f}{f_0}v'^2 -\frac{2}{\sqrt{f_0}} v' z'\right) dx^2,
\end{equation}
where once more we have $'\equiv d/dx$. This now gives
\begin{align}
	&\sqrt{\gamma}=\frac{L^3}{\sqrt{f_0}}\frac{1}{z^3}\left(f_0-fv'^2 -2\sqrt{f_0} v'z'\right)^{1/2}\ , \\
	&\lambda_{GB} L^2\sqrt{\gamma}R_{\Sigma}=(2L^3\lambda_{GB}\sqrt{f_0})\frac{z'^2}{z^3\left(f_0-fv'^2 -2\sqrt{f_0} v'z'\right)^{1/2}} +\frac{dF}{dz}\ ,\label{eq:GH_cancels_dF} 
\end{align}
where
\begin{equation}
	F(x)=(4L^3\lambda_{GB}\sqrt{f_0})\frac{z'}{z^2\left(f_0-fv'^2 -2\sqrt{f_0} v'z'\right)^{1/2}} 
\end{equation}
Clearly, the total derivative term will not contribute to the equations of motion. Furthermore, the $dF/dx$ term is exactly cancelled by the Gibbons-Hawking term\footnote{Note that in \cite{Dong:2013qoa},\cite{Camps:2013zua}  the prescription does not include a boundary term and $dF/dx$ is not cancelled. It is clear that the solutions of the extremization procedure will be the same whether $dF/dx$ is present or not. However, the value of the entanglement entropy might change. In the present case one can check that the contribution of  $dF/dx$ is divergent and thus will be subtracted after normalization.We thank Tom\'as Andrade for comments on this issue}  .  Thus, the effective action that needs to be extremized is given 
\begin{equation}\label{eq:Seff_time}
	S_{eff}=\frac{L^3}{4G_N\sqrt{f_0}}\int \frac{dz}{z^3}\left[\left(f_0-fv'^2 -2\sqrt{f_0} v'z'\right)^{1/2}+\frac{2\lambda_{GB} f_0z'^2}{\left(f_0-fv'^2 -2\sqrt{f_0} v'z'\right)^{1/2}}\right].
\end{equation}

The equations of motion derived from \eqref{eq:Seff_time} are, 
	
\begin{align}\label{eq:eom_ee}
  z^{\prime \prime} &= \frac{F_z(z,z',v,v')}{G(z,z',v,v')}\\
	v^{\prime \prime} &= \frac{F_v(z,z',v,v')}{G(z,z',v,v')}
\end{align}

where we are not writing explicitly the $x$ dependence in $z(x)$, $v(x)$ and $'$ denotes derivative with respect to $x$.

\begin{align}
		F_z &=\sqrt{f_0} f(z,v) \bigg[\sqrt{f_0} v'^2 \left[\left(6 \lambda_{GB} z z'^2-z\right) \partial_zf(z,v)-24 z'^2\right]\\ \nonumber
		&+4 z v'^3 z' \partial_z f(z,v)+z v'^4 \partial_v f(z,v)+6 f_0^{3/2} \left(2 \lambda_{GB}  z'^2-1\right)-24 f_0 v' z' \left(\lambda_{GB}  z'^2-1\right)\bigg]\\ \nonumber
		&+v'^2 f(z,v)^2 \left[z v'^2 f^{(1,0)}(z,v)-24 \sqrt{f_0} v' z'-12 f_0 \left(\lambda_{GB}  z'^2-1\right)\right]\\ \nonumber
		&+f_0 z v' \big[v' \partial_v f(z,v)+2 z' \partial_z f(z,v)\big] \left[ 6 \lambda_{GB} \sqrt{f_0} z'^2-\sqrt{f_0}+2 v' z'\right]-6 v'^4 f(z,v)^3
\end{align}

\begin{align}
F_v &= z v' \partial_z f(z,v) \bigg[-2 \sqrt{f_0} z' \left[v'^2 -4 \lambda_{GB} v'^2 f(z,v)+4 f_0 \lambda_{GB} \right]\\ \nonumber
& +v' \left[f_0-v'^2 f(z,v)\right]+10 f_0 \lambda_{GB}  v' z'^2\bigg]-2 \left[f_0-v' \left(v' f(z,v)+2 \sqrt{f_0} z'\right)\right]\\ \nonumber
&\times \left[2 \sqrt{f_0} \left(\lambda_{GB}  z v'^2 \partial_v f(z,v)+3 z' \left(\sqrt{f_0} \lambda_{GB}  z'+v'\right)\right)+3 v'^2 f(z,v)-3 f_0\right]
\end{align}

\begin{equation}
  G=  2 \sqrt{f_0} z\left[ (4 \lambda_{GB}  f(z,v)+1) \left(v'^2 (-f(z,v))-2 \sqrt{f_0} v' z'+f_0\right)-6 f_0 \lambda_{GB}  z'^2\right]
\end{equation}

We will solve these equations of motion subject to the initial conditions,

\begin{equation}
z(0) = z_*,\qquad \qquad z'(0) = 0
\end{equation}
and,
\begin{equation}
v(0) = v_*, \qquad\qquad 
v'(0) = 0.
\end{equation}

Thus a particular solution is labeled  by $(z_*,v_*)$. We are interested in $(z_*,v_*)$that produce surfaces anchored at the boundary. We also want  relate this constants to boundary quantities $\ell$ and $t_b$. The numerical procedure used to do this is explained below.

\subsection{Numerical procedure}\label{sec:Numerics}

In order to solve the motion equation  numerically, we
employ the Dormand-Prince method \cite{Dormand}, which is an explicit method
to solve systems of differential equations. This method uses six evaluations to compute the fourth
and fifth order solutions and employ them to estimate the relative error, once
determined the error, the program uses an automatic step adjusting procedure.
In the present work we used the C language implementation of the Dormand-
Prince method provided by \cite{Dop853}.
Notice that the condition that the homology condition, in this case equivalent to demanding that the surface is anchored at the boundary, is not included a priori on the system
of differential equations,  they need to be considered during the selection
of initial conditions. Because the solutions diverges at the boundary, it is not
possible to set the initial conditions there. However, since we know that at least
in two points $z\rightarrow 0$, there is a maximum point $z_*$ , by translational invariance
we may set
\begin{equation}
z(0) = z_*,\qquad \qquad z'(0) = 0
\end{equation}

this completely fixes the initial conditions for
$z(x)$. In a similar way we impose the initial conditions for $v(x)$ to be
\begin{equation}
v(0) = v, \qquad\qquad 
v'(0) = 0.
\end{equation}
Then we may parametrize the set of all the solutions by $(z_* , v_*)$.
We want to relate the parameters $(z_*, v_*)$ with the parameters at the bound
ary $(\ell , t_b )$, to do so we notice that, by symmetry, $z(\pm \ell /2) = 0$ and
$v(\pm \ell /2) = t_b $. These equations stablish a relation between the two set of
parameters, however it is not possible to solve them in general. Instead of that
we proceed to scan de parameter space $(z_*, v_*)$ in the region $z_* \in  (0, 10]$ y
$v_* \in  [-10, 10]$ using a grid of$ 600 \times  600$ or $800 \times  800$ over applied over different
subregions. The grid size and the region to be explore were chosen to maxi-
mize the number of curves satisfying the homology condition and crossing the
apparent horizon.
In order to solve the system of differential equation over the grid in a more
efficient way we divide the process into 10 or 20 parallel processors.
Once obtained a list of acceptable initial conditions, those which are anchored at the boundary, we proceed to choose those solutions which satisfied
any required boundary conditions, i.e. $\ell$ and $t_b$ , the precision with which
those values are determined depends on the grid step, for the present work we
keep a precision in the boundary condition determination of $\pm 0.05$
After solving the system of equations and properly imposing the homology
condition we now need to evaluate the entanglement entropy function on-shell,
this procedure needs to be done numerically as well.

We need to be careful due
the divergent behaviour of the integral at the boundary. For the static cases
this is solved by using the conserved quantity to write the integral in terms of
$z$, imposing a UV cut-off $z_0$ and subtracting the divergent contribution. 
In static Gauss-Bonnet backgrounds the  divergent contribution can be computed in a standard way in the static limit by subtracting the vacuum contribution. However, it will depend on $\lambda_{GB}$. This stems from the fact that vacuum Gauss-Bonnet is an AdS space but with a $\lambda_{GB}$ dependent radius. 

In the
dynamic case that procedure cannot be employed since we have no conserved
quantities, although the divergent contribution to the integral is the same as in
the static case, the problem arises when determining the proper way to impose
a UV cut-off in a consistent way, this is which match with the static cut-off
in the asymptotic limit. Instead of directly impose a cut in the x interval, we
employ the numerical integration algorithm itself, imposing a desire precision
and a maximal number of step reduction, once reached those values the integral
is declared as divergent and the integration finished at that value of x.

\subsection{Results}\label{subsec:results}
\begin{figure}
\centering
\includegraphics[width=.45\textwidth,height=.42\textwidth]{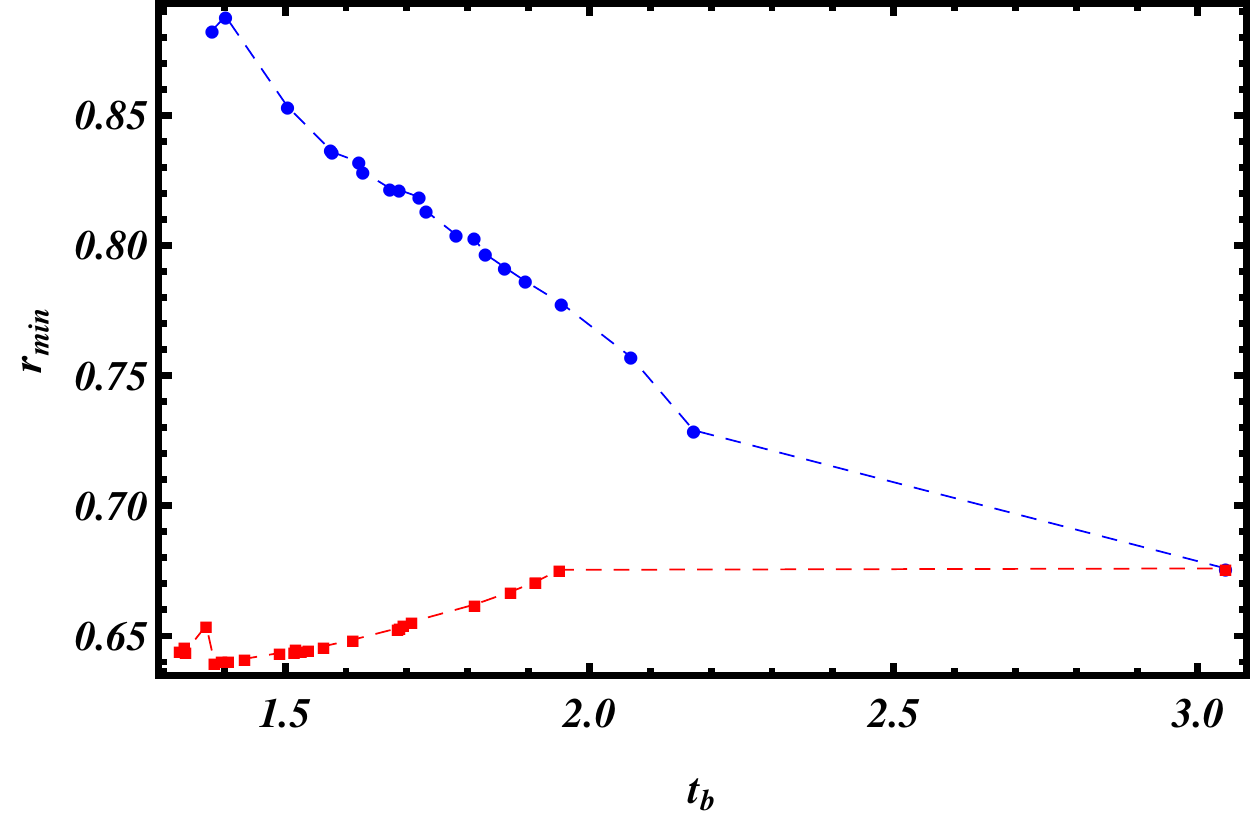}
\includegraphics[width=.45\textwidth, height=.44\textwidth]{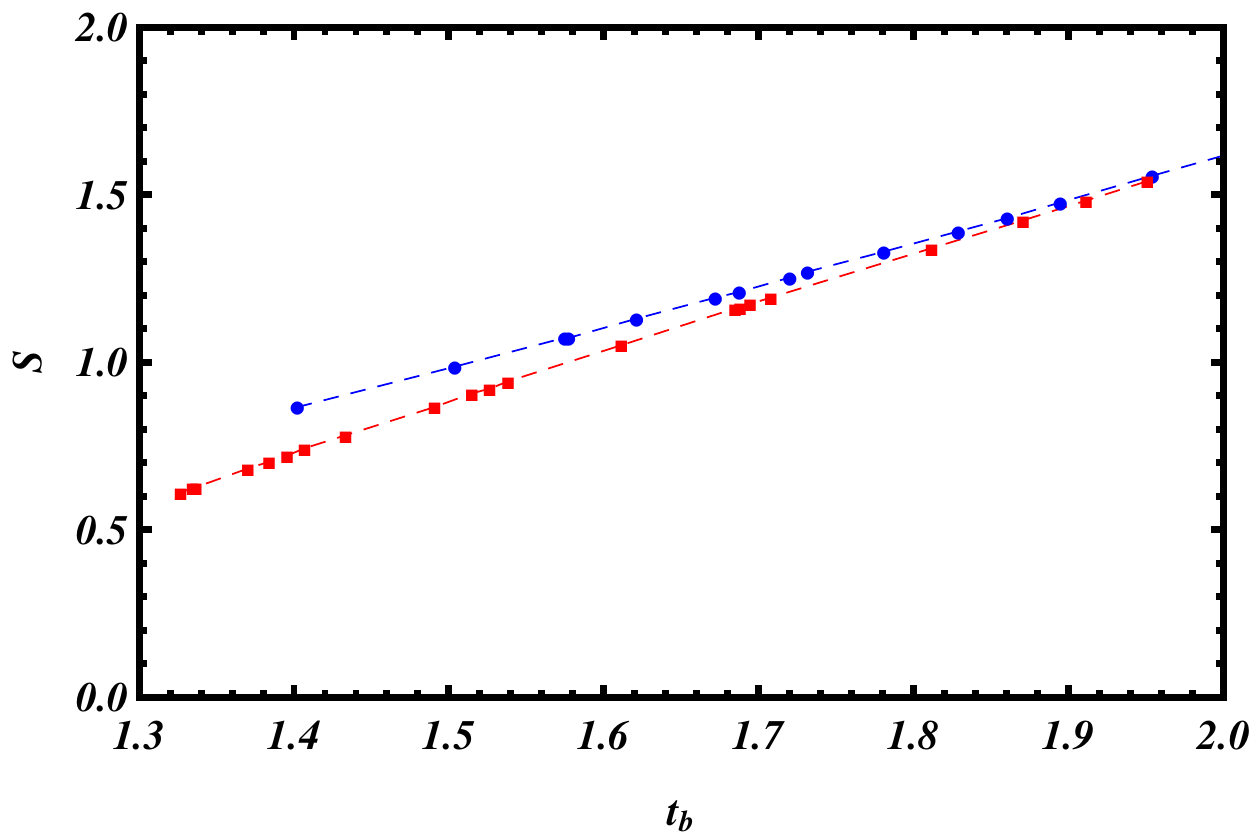}
\caption{\label{fig:min_two_families}Left panel:  minima for the two families of extremal surfaces  $\mathcal{M}^S$, red and $\mathcal{M}^0$, blue. Right panel: Their correponding entanglement entropy .} 
\end{figure}

We carried out a  numerical study of HEE in backgrounds with  $\lambda_{GB}=0.05$ and $\lambda_{GB}=-.05$ as representatives of theories with positive and negative $\lambda_{GB}$. We find that while for positive $\lambda_{GB}$ the behavior of the extremal surface solution of \eqref{eq:eom_ee} is qualitatively similar to the case of Einstein gravity, for negative $\lambda_{GB}$ it is strinkingly different. We summarize our results below:
\begin{itemize}
\item For $\lambda_{GB} <0$ and  for early times and $\ell \ge 2.5$ a new family of solution appears. That is, for a given $t_b$ and $\ell$ two solutions are possible: one that  behaves just like $\lambda_{GB}>0$, we denote this family $\mathcal{M}^0$,  and one that probes arbitrarily close to the singularity, $\mathcal{M}^S$, Fig.\eqref{fig:profile_neg}.
\begin{figure}[tbp]\centering
\includegraphics[width=.45\textwidth ,height=.38\textwidth]{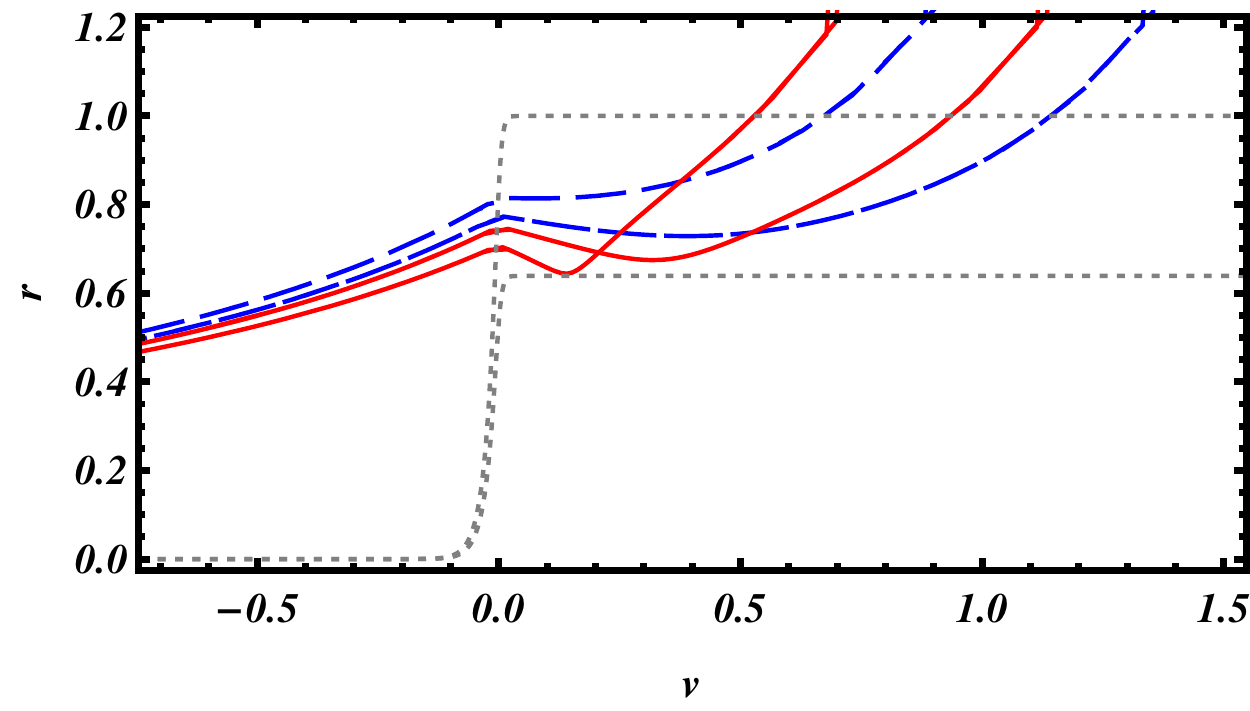}
\caption{Representatives of the two families of minimal surfaces for negative Gauss Bonnet coupling, $\lambda_{GB}=-.05$ . The family $\mathcal{M}^S$ , red curves, contains surfaces that probe arbitrarily close to the singularity. The family $\mathcal{M}^0 $ , dashed curves, is very similar to the $SAdS$ and $\lambda_{GB}>0$, they explore behind the horizon but do not reach the singularity. \label{fig:profile_neg} }

\end{figure}
\item 
We studied the  minima, $r_{min}$, reached by the extremal surfaces and present the results in Fig. \eqref{fig:rmin}. We can see that for early times the family $\mathcal{M}^S$ probes arbitrarily close to the singularity $r_{min}^S \sim r_s$, where $r_s^{-1}$ is defined in \eqref{eq:z_sing}. As time increases $r_{min}^S$ becomes larger and for later times the minima converge such that $r_{min}^S\sim r_{min}^0\sim r_{min}^{\lambda_{GB}>0}\sim r_{min}^{SAdS}$.
\begin{figure}[tbp]
\centering
\includegraphics[width=.5\textwidth]{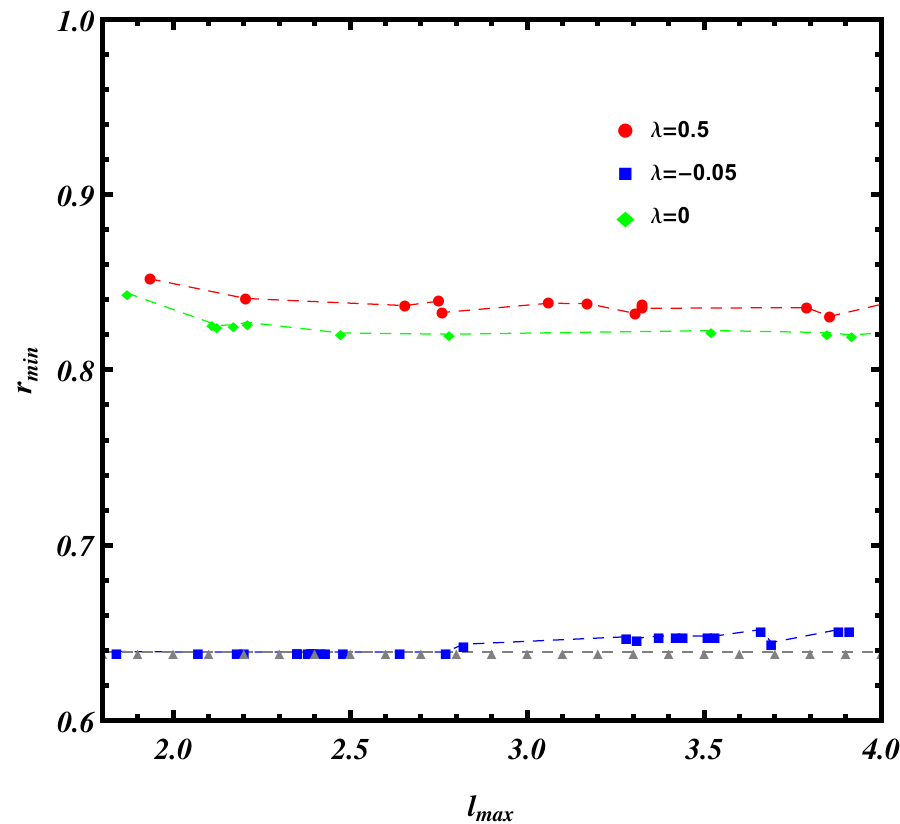}
\includegraphics[width=.5\textwidth, height=.45\textwidth]{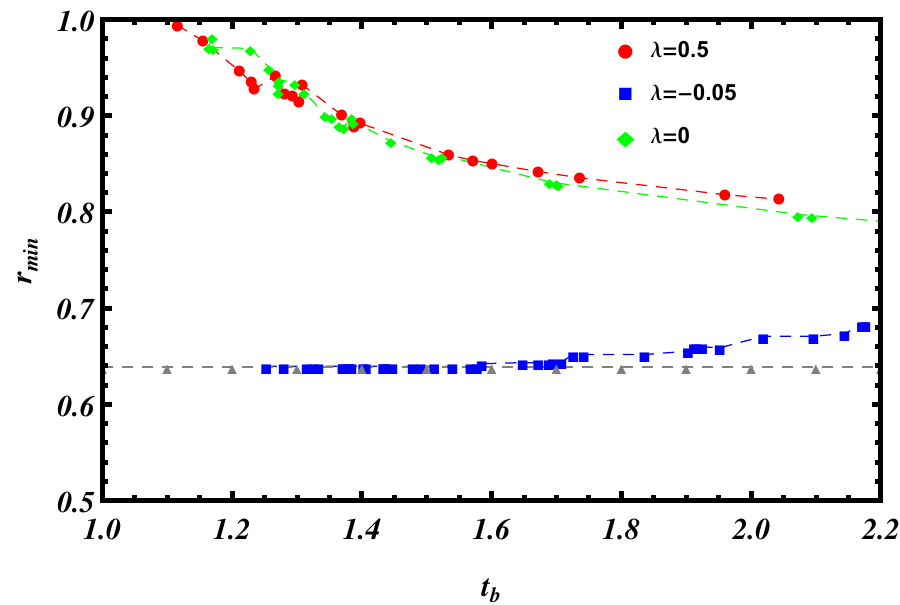}
\caption{
\label{fig:rmin} The minimum point reached by a given  entanglement entropy  surface for $SAdS$, green, $\lambda_{GB}>0$, red, and $\lambda_{GB} <0$, blue. Top panel: $r_{min}$ as function of the boundary separation $\ell$. Bottom panel: $r_{min}$ as function of the boundary time $t_b$. }
\end{figure}

\item We evaluate the functional  in these solutions and find that the one that penetrates deeper behind the horizon, close to the singularity, is the one that represents the entanglement entropy Fig.\eqref{fig:min_two_families}.
\item As a check of our solutions we calculated entanglement entropy as a function of $\ell$. The concavity of this curve is associated with the validity of strong subadditivity \cite{Callan:2012ip},\ \cite{Allais:2011ys} . We find that our solutions obey SSA as expected.
\item The $\lambda_{GB}>0$ case is very similar to $SAdS$, the extremal surfaces penetrate the horizon but only up to a limiting surface. When we compare the minima we find that for $\lambda_{GB}>0$ the extremal surfaces explore less than in the $AdS$ geometry Fig.\eqref{fig:rmin}.
\item As mention in section \eqref{sec:main} the prescription of for the entanglement entropy in a Gauss Bonnet theory is not just a minimal volume as in Einstein gravity. It is natural to ask if the new effects seen here for $\lambda_{GB}<0$ are due to the extra term in the functional or if they are already present in a minimal volume. In order to elucidate this point we calculate the minimal volume for strip regions in an asymptotically AdS Gauss Bonnet black hole \eqref{eq:timemetric}. We find that the minimal volumes also reach close to the singularity in the case of $\lambda_{GB}<0$. We present the corresponding figures in Appendix \eqref{app:volume}.

\end{itemize}

\begin{figure}
\centering
\includegraphics[width=.45\textwidth,height=.4\textwidth]{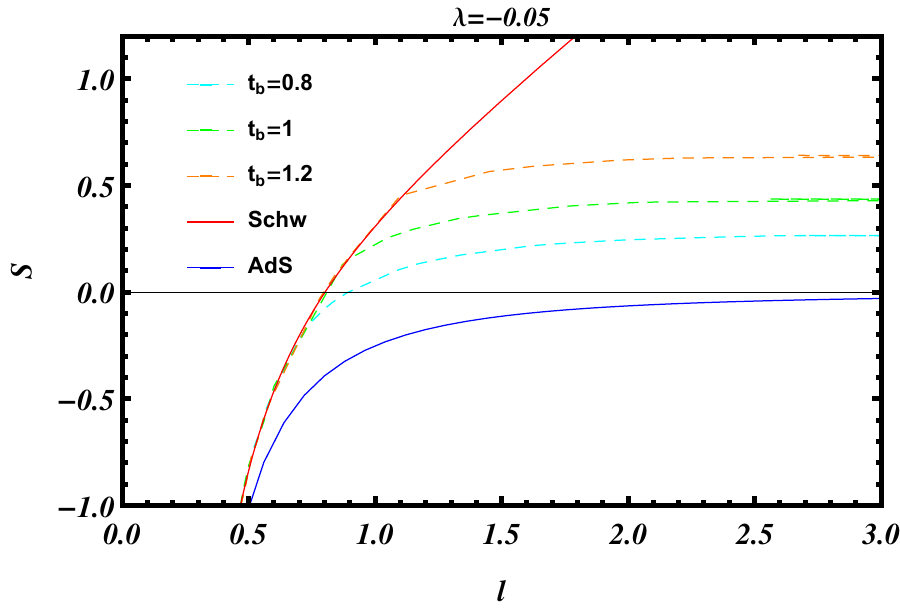}
~
\includegraphics[width=.45\textwidth,height=.4\textwidth]{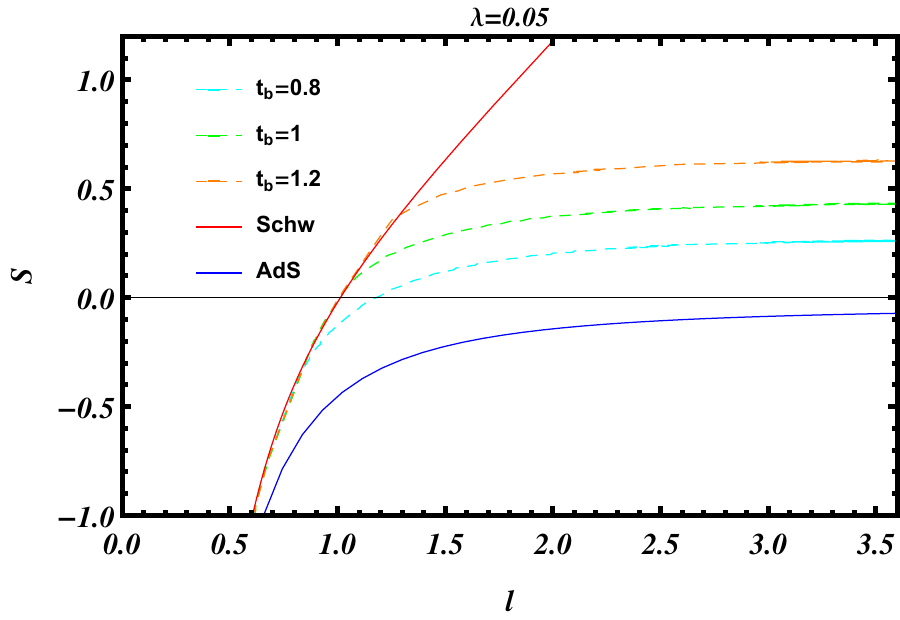}
\caption{\label{fig:SSA} Entanglement entropy as a function of $\ell$ for $\lambda_{GB} =-.05 $ (left panel) and $\lambda_{GB}=.05$ (right panel)}

\end{figure}

\section{Conclusions and future directions}\label{sec:conclusions}

The main motivation of this work is a simple question,  {\it how far behind the horizon does the HEE probe reach in Gauss-Bonnet theories?}. Much is known about the similar question in Einstein gravity but previous studies in Gauss-Bonnet have focused in finding the thermalization time and  not in the reach of the probes. In the AdS/CFT context higher derivative theories are interesting because they are  dual to field theories with corrections in $1/\sqrt{g_{YM} N}$. We chose  Gauss-Bonnet as example of a higher derivative theory because of its solvability; there are black hole solutions known analytically.  We take Gauss-Bonnet as a toy model where to learn features of holographic entanglement entropy in a higher derivative theory and not as a dual of a particular field theory. As a warmup we studied geodesics in a Vaidya Gauss-Bonnet background.  When $\lambda_{GB}$ is positive there are no surprises, the holographic probes behaves as in Einstein gravity. Namely, they  penetrate behind the apparent horizon but not arbitrarily close to the singularity; they asymptote to a certain critical surface. A different and  novel behavior appears when we consider $\lambda_{GB}<0$: the solutions become double valued. A new family of solutions that  can reach arbitrarily close to the singularity appears. We observe the same behavior in minimal volume surfaces. However, in a 5 dimensional bulk these objects are purely geometric, they are not dual to any field theory observable. 

Unlike in Einstein gravity, the holographic entanglement entropy prescription for higher derivative theories \cite{Camps:2013zua},\cite{Dong:2013qoa} is not only a minimal area but includes an extra term involving the extrinsic curvature. We study the entanglement entropy in a Vaidya Gauss-Bonnet background and quantify how much behind the horizon the probes reach, $r_{min}$ , and find that for $\lambda_{GB}> 0$ the probes explore less than in Einstein gravity, {\it i.e} $r_{min}^{GB}> r_{min}^{Einstein}$. 

For $\lambda_{GB}<0$ we find two family of solutions: one that behaves very similar to Einstein gravity and a new one that reaches arbitrarily close to the singularity. 
Having two solutions the HRT prescription instruct us to choose the one of minimal entropy. We find that the surfaces reaching the singularity are the ones of minimal entropy.  Thus, {\it for $\lambda_{GB} <0$ the holographic entanglement entropy can probe all the way to the singularity}. This is the main result of the present work. As a check of our solutions  we verify the concavity of $S(l)$ impliying that SSA is respected.

Let us point some open problems and future directions related to the present work;
\begin{itemize}
\item {\it Lionhearted effort:} Because of its time dependent nature, the problem studied here is numerically  intensive. We have concentrated in $\lambda_{GB}=0.05$ and $\lambda_{GB}=-0.05$ as representatives of positive and negative couplings. A complete analysis scanning over  a range of values of $\lambda_{GB}$ would certainly  be desirable and interesting and might uncover interesting physics as $\lambda_{GB}$ becomes larger. In particular, the novel solutions discovered here  for negative $\lambda_{GB}$ (the ones that can reach the singularity) exist only for a very short time after the probe has crossed the shell. It is natural to think that as $\lambda_{GB}$ grows more negative these type of solutions will exist for a larger time. Or it could be that the opposite is true, that  for larger negative $\lambda_{GB}$ these solutions cease to exist. The only way to answer these questions is a full numerical  analysis over the $\lambda_{GB}$ parameter space. 

\item  An inmediate generalization of the present work would be to study spacetimes of dimensions higher than 5 and  boundary regions other than the strip. 

\item It would be interesting to ask the same question investigated here in other higher derivative theories like more general Lovelock theories where many black hole solutions are known \cite{Camanho:2011rj}. Also, thermalization in hyperscaling violating backgrounds was investigated in \cite{Fonda:2014ula},\cite{Alishahiha:2014cwa} and higher derivative corrections in  \cite{Bueno:2014oua},\cite{Knodel:2013fua}. Thus, extending the present work to hyperscaling violating geometries seems a viable and interesting endeavor.

\item We have studied entanglement entropy in a time dependent Gauss-Bonnet background in  Poincare patch. It the static case, it is known that  novel features of the entanglement entropy  appear when considering compact spaces \cite{Hubeny:2013gta}. Thus, extending  the present work to  spherically symmetric spaces might also uncover some new phenomena sensitive to the sign of the Gauss-Bonnet coupling $\lambda_{GB}$.

\item 
In the same spirit of understanding how much of the bulk does the HEE probes,  it would be interesting to consider an  asymptotically global AdS static black hole in Gauss-Bonnet gravity and  study the  entanglement shadow. We expect that entanglement  shadow will increase or decrease (as compared to Einstein gravity) depending on the sign of $\lambda_{GB}$.

\item It would be interesting to perform as similar  study with different  holographic probes like the causal holographic information. The acausality of the boundary theory for finite  $\lambda_{GB}$ might be reflected in some particular behavior of the causal holographic information surface $\chi_A$.

\item In \cite{Bhattacharyya:2009uu} a formalism was developed to study black hole formation in a weak field limit. As shown in \cite{Caceres:2014pda} some of the interesting physics discovered using the  Vaidya model for charged black holes \cite{Caceres:2012em}   can be captured in the weak  field approach. Although valid only after a certain time after collapse, the advantage of the perturbative approach is that it is numerically simpler. Thus, it  would be interesting to investigate  scalar collapse in a Gauss Bonnet theory in the weak field limit and see if some of the results presented here can be also obtained in that framework.
\end{itemize}

We hope to  return to  some of these problems in the near future.

\section*{Acknowledgments}
It is a pleasure to thank Tom\'as Andrade, Veronika Hubeny, Cindy Keeler,  Yi Pang and Juan Pedraza for useful discussions. 
This research was supported  by  Mexico's National Council of Science
and Technology (CONACyT) grant CB-2014-01-238734 and  the National Science Foundation under Grant PHY-1316033 and Grant  No. NSF PHY11-25915.
E.C  thanks the Galileo Galilei Institute for Theoretical Physics and the Kavli Institute for Theoretical Physics  for  hospitality and the INFN for partial support during the completion of this work.

\appendix
\section{Gibbons-Hawking term}
In this appendix we provide details of the calculation of the boundary term,
\[
S_{GH}=\oint_{\partial \Sigma}\sqrt{h}  \mathcal{K}.
\]\label{eq:bdyterm}
To fix notation, recall that the integral \ref{eq:bdyterm}  is over the boundary of the codimension 2 surface $\Sigma$ with induced metric $\gamma$,  $h$ is the metric induced at the boundary and $\mathcal{K}$ the trace of the extrinsic curvature of $\Sigma$. 

As we saw in \ref{eq:codimtwo}, the metric induced in the co-dimension two surface is
\begin{equation}\label{eq:codimtwo_app}
	\gamma_{ab} dx^a dx^b  = \frac{L^2}{z^2}\left(1- \frac{f}{f_0}v'^2 -\frac{2}{\sqrt{f_0}} v' z'\right) dx^2+\frac{L^2}{z^2}(dx_2^2 + dx_3^2),
\end{equation}
The unit norm vector perpendicular to the boundary is  clearly in the $x$ direction,
\[\eta=(\frac{L}{z}\sqrt{\left(1- \frac{f}{f_0}v'^2 -\frac{2}{\sqrt{f_0}} v' z'\right)}  ,0,0)\]
the trace of the extrinsic curvature is then
\[ \mathcal{K}= \gamma^{ab} \nabla_a\eta_b = 2 \frac{z' \sqrt{f_0}}{L\sqrt{f_0+f v'^2 -2 \sqrt{f_0}v'z'}}\]
Now, the determinant of the metric induced at the boundary is simply $\frac{L^2}{z^2}$. Thus we have 
\[S_{GH}=\oint_{\partial\gamma} dx_2 dx_3  2 L \frac{z' \sqrt{f_0}}{ z^2 \sqrt{f_0+f v'^2 -2 \sqrt{f_0}v'z'}}\]
which exactly cancels the term $\int \frac{dF}{dx}$ in \ref{eq:GH_cancels_dF}.

\section{Equations of motion}\label{app:eom}

\subsection{Minimal volume}

\begin{equation}
\mathcal{L} = \frac{1}{z^3}\sqrt{f_0 - f(z, v) v'^2 -2 \sqrt{f_0} v' z'} 
\end{equation}

The equations of motion are,
	
\begin{align}
  z^{\prime \prime} &= \frac{F_z(z,z',v,v')}{G(z,z',v,v')}\\
	v^{\prime \prime} &= \frac{F_v(z,z',v,v')}{G(z,z',v,v')}
\end{align}
where we are not writing explicitly the $x$ dependence in $z(x)$, $v(x)$ and $'$ denotes derivative with respect to $x$.

\begin{align}
		F_z &=-(6f(z,v)^2 v'^2 + f(z,v)(6f_0+12\sqrt{f_0}v'z'+ z v'^2 \partial_z f(z,v))\\ \nonumber
&\sqrt{f_0}z v'( v'\partial_v f(z,v) + 2 z'\partial_z f(z,v)))	
	\end{align}

\begin{align}
F_v &= 6 f_0 - 12 \sqrt{f_0} v'z' + v'^2( -6 f(z,v)+ z \partial_zf(z,v))
\end{align}

\begin{equation}
  G=  2 \sqrt{f_0} z
  \end{equation}

\subsection{Geodesics} 

\begin{equation}
\mathcal{L} = \frac{1}{z}\sqrt{f_0 - f(z, v) v'^2 -2 \sqrt{f_0} v' z'} 
\end{equation}

The equations of motion are,
	
\begin{align}
  z^{\prime \prime} &= \frac{F_z(z,z',v,v')}{G(z,z',v,v')}\\
	v^{\prime \prime} &= \frac{F_v(z,z',v,v')}{G(z,z',v,v')}
\end{align}
\begin{align}
		F_z &=-(2f(z,v)^2 v'^2 + f(z,v)(2 f_0+ 4\sqrt{f_0}v'z'+ z v'^2 \partial_z f(z,v))\\ \nonumber
&\sqrt{f_0}z v'( v'\partial_v f(z,v) + 2 z'\partial_z f(z,v)))	
	\end{align}

\begin{align}
F_v &= 2 f_0 - 4 \sqrt{f_0} v'z' + v'^2( -2 f(z,v)+ z \partial_zf(z,v))
\end{align}

\begin{equation}
  G=  2 \sqrt{f_0} z
  \end{equation}

\section{Minimal Volume results}\label{app:volume}
\begin{figure}
\centering
\parbox{3in}{\includegraphics[width=.45\textwidth]{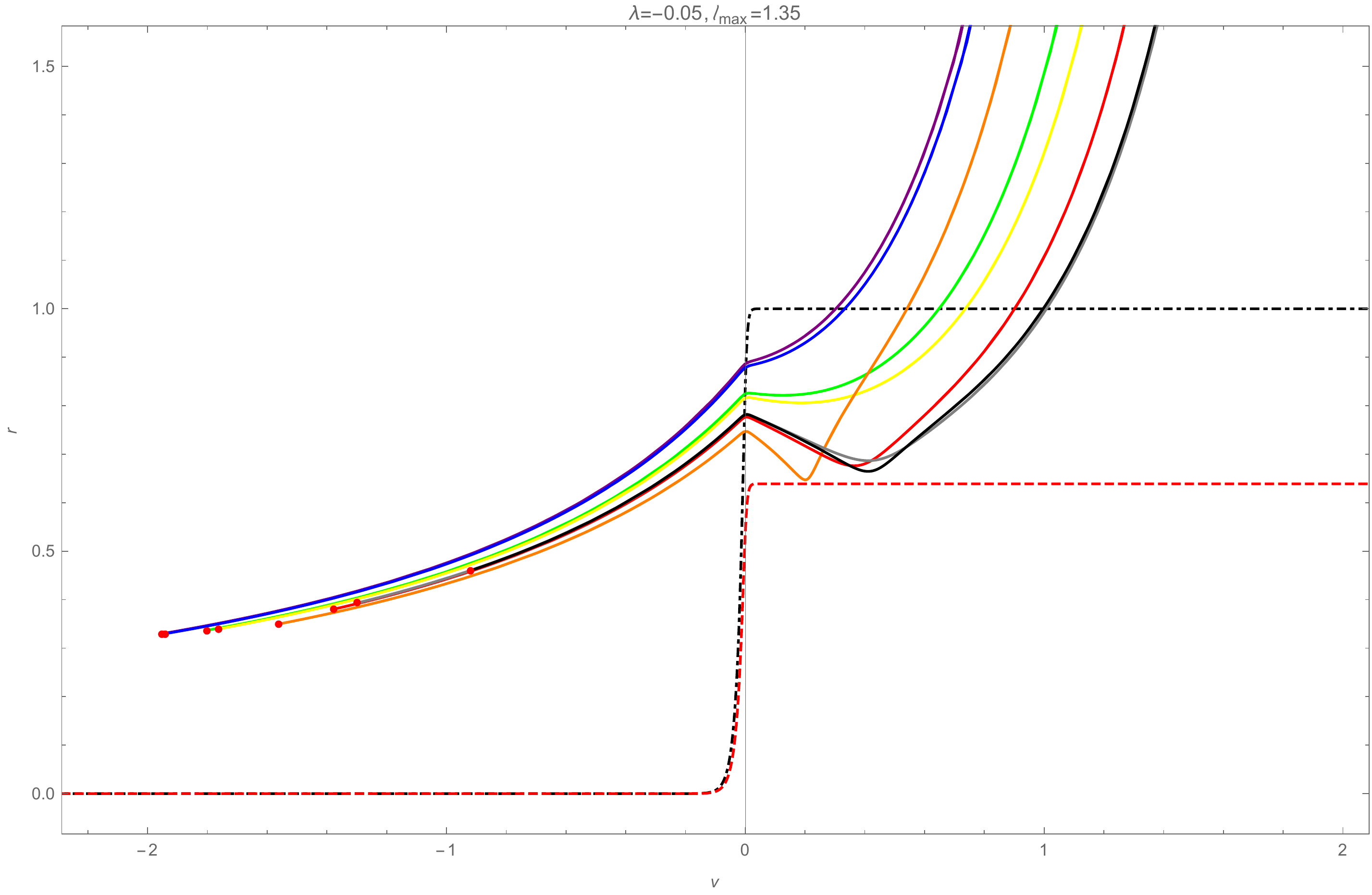} } 
\parbox{3in}{\includegraphics[width=.45\textwidth]{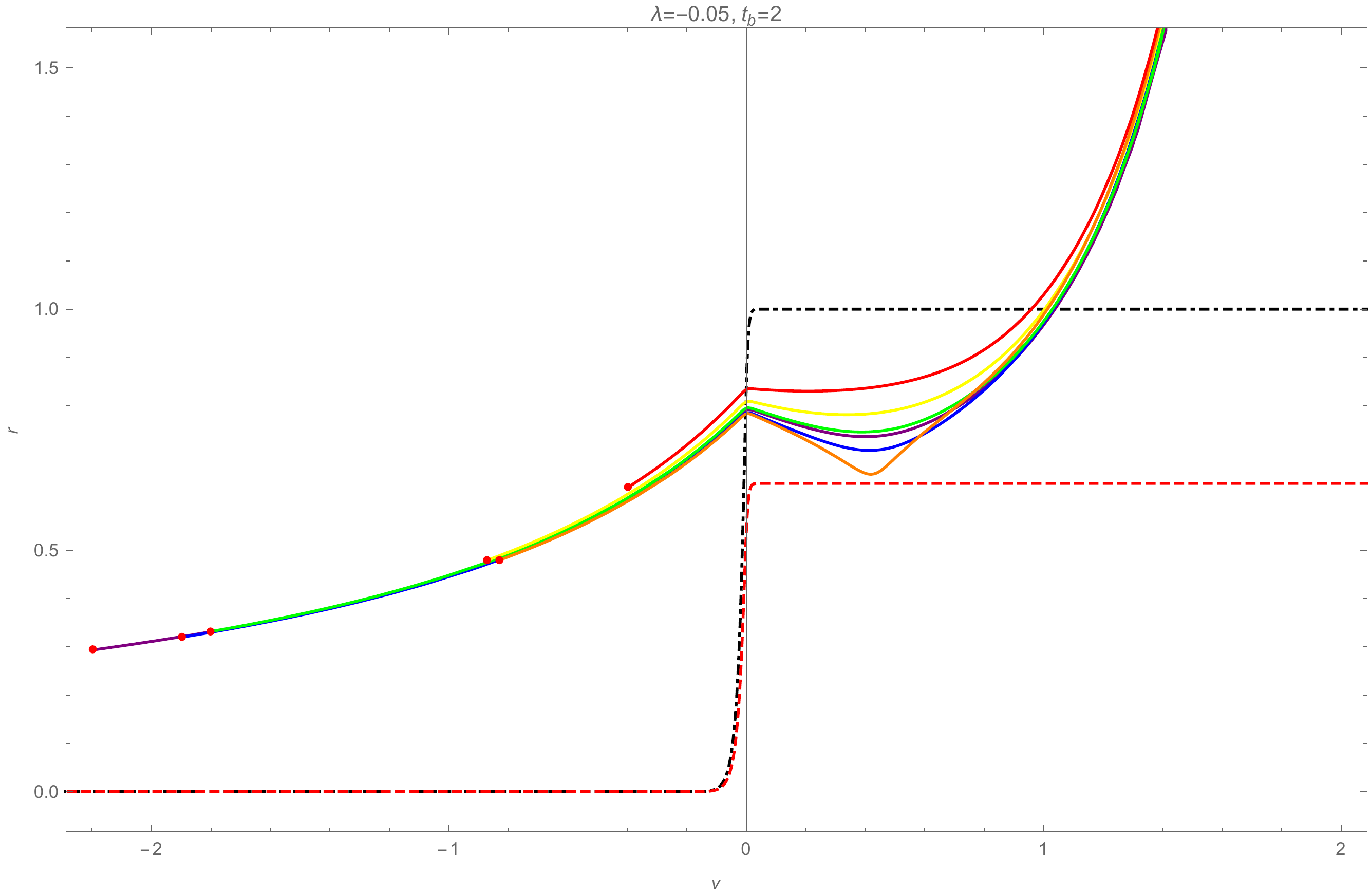}  }
\caption{Left panel: $r$ vs $v$ profile of representative geodesics with $\lambda_{GB}=-0.05$ and fixed $\ell=1.35$. Right panel: $r$ vs $v$ profile of representative geodesics with $\lambda_{GB}=-0.05$ and fixed $t_b= 2$}
\label{fig:profiles_neg_vol_pretty}
\end{figure}

\begin{figure}
\centering
\parbox{3in}{\includegraphics[width=.45\textwidth]{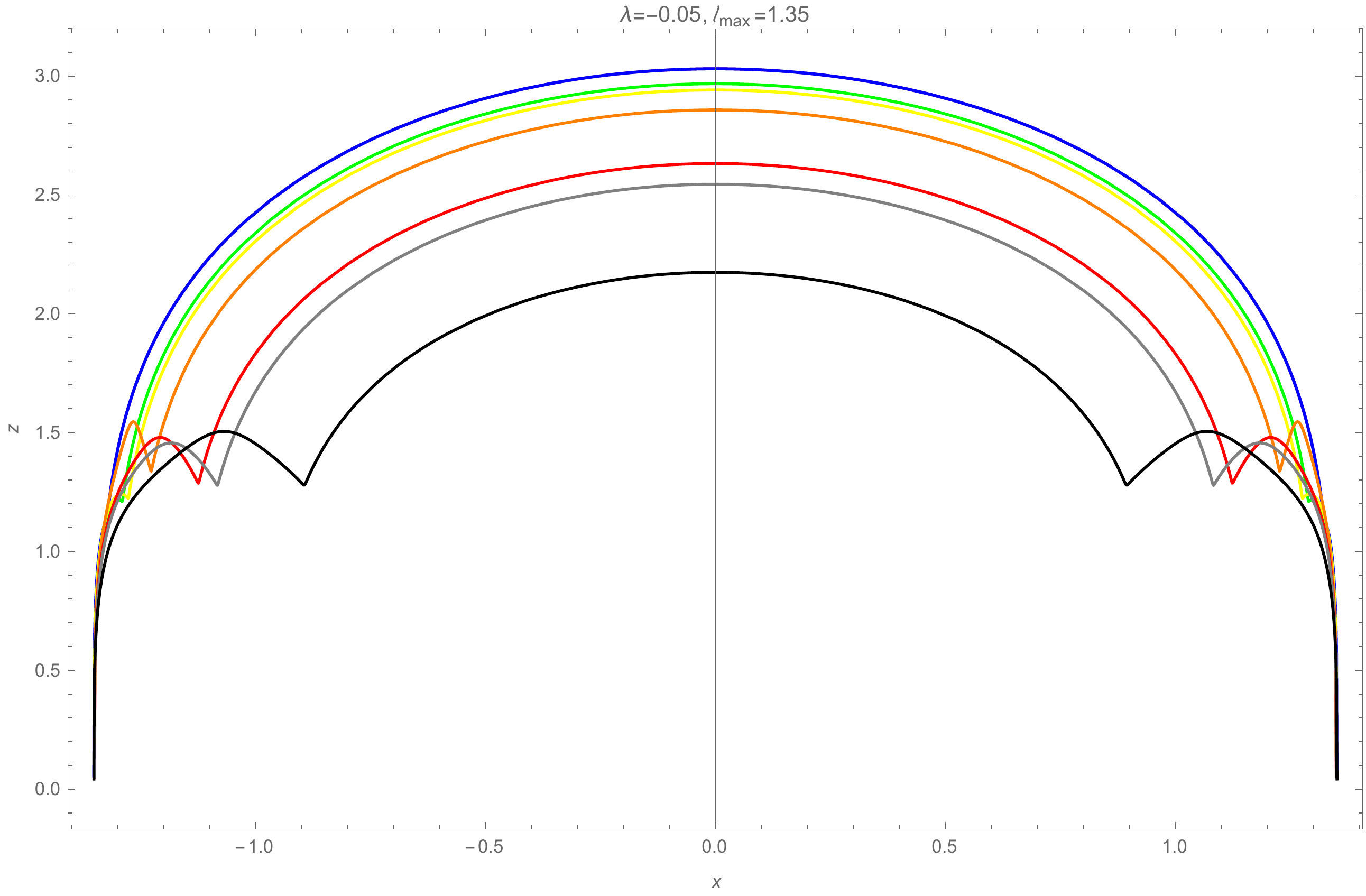} }
\parbox{3in}{\includegraphics[width=.45\textwidth]{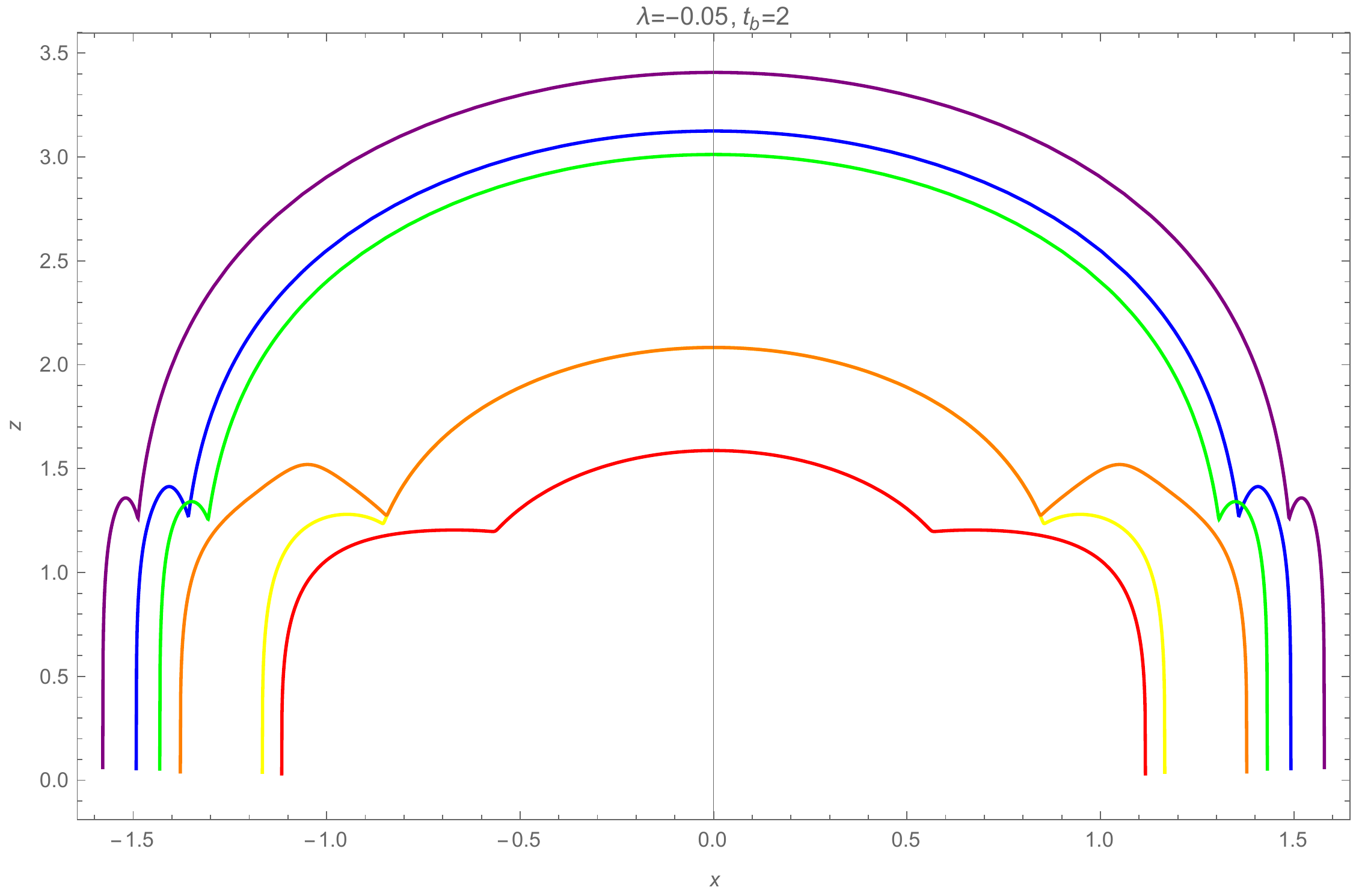} }
\caption{Left panel:$x$ vs $z$ profile of representative geodesics with $\lambda_{GB}=-0.05$ and fixed $\ell=1.35$. Right panel: $x$ vs $z$ profile of representative geodesics with $\lambda=-0.05$ and fixed $t_b= 2$}
\label{fig:profiles_vol_fat}
\end{figure}

In order to understand if the effects found for $\lambda_{GB}$ in the entanglement entropy are due to the correction in the prescription for \eqref{eq:ee} or are inherent to minimal surfaces in Gauss-Bonnet we study  volumes in the background \eqref{eq:timemetric}. These are purely geometrical objects that are not dual to any observable in the field theory. We find, Figs.\eqref{fig:profiles_neg_vol_pretty}, \eqref{fig:profiles_vol_fat},  that for $\lambda_{GB} <0$ there are minimal volume surfaces that, for early times,  reach the singularity. 

\bibliography{HD}{}
\bibliographystyle{JHEP}
\end{document}